%% file: main.tex
\pdfoutput=1
\documentclass[12pt,a4paper]{article}
\usepackage{ifthen} % for conditional statements
\newboolean{pdflatex}
\setboolean{pdflatex}{true} % False for eps figures 

\newboolean{articletitles}
\setboolean{articletitles}{true} % False removes titles in references

\newboolean{uprightparticles}
\setboolean{uprightparticles}{false} %True for upright particle symbols

\usepackage{booktabs}
\usepackage{multirow}
\usepackage{adjustbox}
\usepackage{lscape}

\def\paperauthors{LHCb collaboration} % Leave as is for PAPER, CONF and FIGURE
\def\paperasciititle{Measurement of the shape of the Bs->Dsstarmunu differential decay rate} % Set ASCII title here
\def\papertitle{Measurement of the shape \break of the $B_s^0\rightarrow D_s^{*-}\mu^{+}\nu_{\mu}$ \break differential decay rate} % Latex formatted title
\def\paperkeywords{{High Energy Physics}, {LHCb}} % Comma separated list
\def\papercopyright{\the\year\ CERN for the benefit of the LHCb collaboration} % new since 9/Apr/2018
\def\paperlicence{CC BY 4.0 licence}
\def\paperlicenceurl{https://creativecommons.org/licenses/by/4.0/}

\input{preamble}
\input{local-symbols-def}

\usepackage{longtable} % only for template; not usually to be used in PAPERs

\begin{document}

%%%%%%%%%%%%%%%%%%%%%%%%%
%%%%% Title     %%%%%%%%%
%%%%%%%%%%%%%%%%%%%%%%%%%
\renewcommand{\thefootnote}{\fnsymbol{footnote}}
\setcounter{footnote}{1}

% %%%%%%% CHOOSE TITLE PAGE--------
%\onecolumn
%\input{title-LHCb-INT}
%\input{title-LHCb-ANA}
%\input{title-LHCb-CONF}
%\input{title-LHCb-FIGURE}
\input{title-LHCb-PAPER}
%\twocolumn
% %%%%%%%%%%%%% ---------

\renewcommand{\thefootnote}{\arabic{footnote}}
\setcounter{footnote}{0}

%%%%%%%%%%%%%%%%%%%%%%%%%%%%%%%%
%%%%%  Table of Content   %%%%%%
%%%%%%%%%%%%%%%%%%%%%%%%%%%%%%%%
%%%% Uncomment next 2 lines if desired
%\tableofcontents
%\cleardoublepage

%%%%%%%%%%%%%%%%%%%%%%%%%
%%%%% Main text %%%%%%%%%
%%%%%%%%%%%%%%%%%%%%%%%%%

\pagestyle{plain} % restore page numbers for the main text
\setcounter{page}{1}
\pagenumbering{arabic}

%% Uncomment during review phase. 
%% Comment before a final submission.
%\linenumbers

% You can include short sections directly in the main tex file.
% However, for larger papers it is desirable to split the text into
% several semiautonomous files, which can be revised independently.
% This is especially useful when developing a document in
% collaboration with several people, since then different parts can be
% edited independently.  This type of file organization is shown here.
% 

\input{introduction}

\input{detector}

\input{selection}

\input{yields}

\input{efficiency}

\input{unfolding}

\input{systematic}

\input{FF}

\input{conclusions}

%\clearpage
% Do not include this in any draft (just for information in the template)
\input{acknowledgements}

% Comment this in for paper drafts; do not include this in analysis note, conference and figure reports

\clearpage
\input{appendix}
\clearpage

\addcontentsline{toc}{section}{References}
%\setboolean{inbibliography}{true}
\bibliographystyle{LHCb}
\bibliography{main,standard,LHCb-PAPER,LHCb-CONF,LHCb-DP,LHCb-TDR}

\newpage
\input{LHCb_Authorship_17-Dec-2019}

%The author list for journal publications is generated from the
%Membership Database shortly after 'approval to go to paper' has been
%given.  It will be sent to you by email shortly after a paper number
%has been assigned.  The author list should be included in the draft used for 
%first and second circulation, to allow new members of the collaboration to verify
%that they have been included correctly. Occasionally a misspelled
%name is corrected, or associated institutions become full members.
%Therefore an updated author list will be sent to you after the final
%EB review of the paper.  In case line numbering doesn't work well
%after including the authorlist, try moving the \verb!\bigskip! after
%the last author to a separate line.

%The authorship for Conference Reports should be ``The LHCb
%collaboration'', with a footnote giving the name(s) of the contact
%author(s), but without the full list of collaboration names.

%The authorship for Figure Reports should be ``The LHCb
%collaboration'', with no contact author and without the full list 
%of collaboration names.

\end{document}

%% file: preamble.tex
\usepackage[top=1in, bottom=1.25in, left=1in, right=1in]{geometry}

\usepackage{microtype}
\usepackage{lineno}  % for line numbering during review
\usepackage{xspace} % To avoid problems with missing or double spaces after
\usepackage{caption} %these three command get the figure and table captions automatically small

\usepackage{graphicx}  % to include figures (can also use other packages)
\usepackage{color}
\usepackage{colortbl}
\graphicspath{{./figs/}} % Make Latex search fig subdir for figures
\usepackage{amsmath} % Adds a large collection of math symbols
\usepackage{amssymb}
\usepackage{amsfonts}
\usepackage{upgreek} % Adds in support for greek letters in roman typeset

\newcommand*\patchAmsMathEnvironmentForLineno[1]{%
\expandafter\let\csname old#1\expandafter\endcsname\csname #1\endcsname
\expandafter\let\csname oldend#1\expandafter\endcsname\csname
end#1\endcsname
 \renewenvironment{#1}%
   {\linenomath\csname old#1\endcsname}%
   {\csname oldend#1\endcsname\endlinenomath}%
}
\newcommand*\patchBothAmsMathEnvironmentsForLineno[1]{%
  \patchAmsMathEnvironmentForLineno{#1}%
  \patchAmsMathEnvironmentForLineno{#1*}%
}
\AtBeginDocument{%
\patchBothAmsMathEnvironmentsForLineno{equation}%
\patchBothAmsMathEnvironmentsForLineno{align}%
\patchBothAmsMathEnvironmentsForLineno{flalign}%
\patchBothAmsMathEnvironmentsForLineno{alignat}%
\patchBothAmsMathEnvironmentsForLineno{gather}%
\patchBothAmsMathEnvironmentsForLineno{multline}%
\patchBothAmsMathEnvironmentsForLineno{eqnarray}%
}

\usepackage{hyperxmp}

\usepackage[colorinlistoftodos,textsize=scriptsize]{todonotes}

\usepackage[bottom,flushmargin,hang,multiple]{footmisc}

\usepackage[pdftex,
            pdfauthor={\paperauthors},
            pdftitle={\paperasciititle},
            pdfkeywords={\paperkeywords},
            pdfcopyright={Copyright (C) \papercopyright},
            pdflicenseurl={\paperlicenceurl}]{hyperref}
\usepackage[all]{hypcap} % Internal hyperlinks to floats.

\input{lhcb-symbols-def} % Add in the predefined LHCb symbols

\usepackage{cite} % Allows for ranges in citations
\usepackage{mciteplus}

%% file: lhcb-symbols-def.tex
\usepackage{xspace} 
\usepackage{upgreek}

\def\lhcb   {\mbox{LHCb}\xspace}

\def\MagUp {\mbox{\em Mag\kern -0.05em Up}\xspace}

\ifthenelse{\boolean{uprightparticles}}%
{
 
 \def\Pgamma      {\ensuremath{\upgamma}\xspace}

 \def\Pmu         {\ensuremath{\upmu}\xspace}                 
 \def\Pnu         {\ensuremath{\upnu}\xspace}                 
                  
 \def\Ppi         {\ensuremath{\uppi}\xspace}                 
                  
 \def\Prho        {\ensuremath{\uprho}\xspace}                 
                  
 \def\Ptau        {\ensuremath{\uptau}\xspace}                 
                  
 \def\Pphi        {\ensuremath{\upphi}\xspace}

 \def\Ppsi        {\ensuremath{\uppsi}\xspace}

 \def\PDelta      {\ensuremath{\Delta}\xspace}                 
 \def\PXi         {\ensuremath{\Xi}\xspace}                 
 \def\PLambda     {\ensuremath{\Lambda}\xspace}                 
 \def\PSigma      {\ensuremath{\Sigma}\xspace}                 
 \def\POmega      {\ensuremath{\Omega}\xspace}                 
 \def\PUpsilon    {\ensuremath{\Upsilon}\xspace}

 \def\PB      {\ensuremath{\mathrm{B}}\xspace}                 
                  
 \def\PD      {\ensuremath{\mathrm{D}}\xspace}

 \def\PJ      {\ensuremath{\mathrm{J}}\xspace}                 
 \def\PK      {\ensuremath{\mathrm{K}}\xspace}

 \def\PW      {\ensuremath{\mathrm{W}}\xspace}

 \def\Pb      {\ensuremath{\mathrm{b}}\xspace}                 
 \def\Pc      {\ensuremath{\mathrm{c}}\xspace}                 
 \def\Pd      {\ensuremath{\mathrm{d}}\xspace}

 \def\Pi      {\ensuremath{\mathrm{i}}\xspace}

 \def\Pp      {\ensuremath{\mathrm{p}}\xspace}

 \def\Ps      {\ensuremath{\mathrm{s}}\xspace}                 
                  
 \def\Pu      {\ensuremath{\mathrm{u}}\xspace}

 \def\thebaroffset{0.0em}
}
{
 
 \def\Pgamma      {\ensuremath{\gamma}\xspace}

 \def\Pmu         {\ensuremath{\mu}\xspace}                 
 \def\Pnu         {\ensuremath{\nu}\xspace}                 
                  
 \def\Ppi         {\ensuremath{\pi}\xspace}                 
                  
 \def\Prho        {\ensuremath{\rho}\xspace}                 
                  
 \def\Ptau        {\ensuremath{\tau}\xspace}                 
                  
 \def\Pphi        {\ensuremath{\phi}\xspace}

 \def\Ppsi        {\ensuremath{\psi}\xspace}                 
                  
 \mathchardef\PDelta="7101
 \mathchardef\PXi="7104
 \mathchardef\PLambda="7103
 \mathchardef\PSigma="7106
 \mathchardef\POmega="710A
 \mathchardef\PUpsilon="7107
                  
 \def\PB      {\ensuremath{B}\xspace}                 
                  
 \def\PD      {\ensuremath{D}\xspace}

 \def\PJ      {\ensuremath{J}\xspace}                 
 \def\PK      {\ensuremath{K}\xspace}

 \def\PW      {\ensuremath{W}\xspace}

 \def\Pb      {\ensuremath{b}\xspace}                 
 \def\Pc      {\ensuremath{c}\xspace}                 
 \def\Pd      {\ensuremath{d}\xspace}

 \def\Pi      {\ensuremath{i}\xspace}

 \def\Pp      {\ensuremath{p}\xspace}

 \def\Ps      {\ensuremath{s}\xspace}                 
                  
 \def\Pu      {\ensuremath{u}\xspace}

 \def\thebaroffset{0.18em}
}
\newcommand{\offsetoverline}[2][\thebaroffset]{\kern #1\overline{\kern -#1 #2}}%

\makeatletter
\ifcase \@ptsize \relax% 10pt
  \newcommand{\miniscule}{\@setfontsize\miniscule{4}{5}}% \tiny: 5/6
\or% 11pt
  \newcommand{\miniscule}{\@setfontsize\miniscule{5}{6}}% \tiny: 6/7
\or% 12pt
  \newcommand{\miniscule}{\@setfontsize\miniscule{5}{6}}% \tiny: 6/7
\fi
\makeatother

\DeclareRobustCommand{\optbar}[1]{\shortstack{{\miniscule (\rule[.5ex]{1.25em}{.18mm})}
  \\ [-.7ex] $#1$}}

\def\mup        {{\ensuremath{\Pmu^+}}\xspace}
\def\mun        {{\ensuremath{\Pmu^-}}\xspace} % muon negative (\mum is taken)

\def\taup       {{\ensuremath{\Ptau^+}}\xspace}
\def\taum       {{\ensuremath{\Ptau^-}}\xspace}

\def\ellp       {{\ensuremath{\ell^+}}\xspace}

\def\neu        {{\ensuremath{\Pnu}}\xspace}
\def\neub       {{\ensuremath{\overline{\Pnu}}}\xspace}
\def\neum       {{\ensuremath{\neu_\mu}}\xspace}
\def\neumb      {{\ensuremath{\neub_\mu}}\xspace}
\def\neut       {{\ensuremath{\neu_\tau}}\xspace}

\def\neul       {{\ensuremath{\neu_\ell}}\xspace}

\def\g      {{\ensuremath{\Pgamma}}\xspace}

\def\W      {{\ensuremath{\PW}}\xspace}

\def\uquark    {{\ensuremath{\Pu}}\xspace}

\def\dquark    {{\ensuremath{\Pd}}\xspace}

\def\squark    {{\ensuremath{\Ps}}\xspace}

\def\cquark    {{\ensuremath{\Pc}}\xspace}

\def\bquark    {{\ensuremath{\Pb}}\xspace}

\def\pion   {{\ensuremath{\Ppi}}\xspace}
\def\piz    {{\ensuremath{\pion^0}}\xspace}
\def\pip    {{\ensuremath{\pion^+}}\xspace}
\def\pim    {{\ensuremath{\pion^-}}\xspace}

\def\rhomeson {{\ensuremath{\Prho}}\xspace}

\def\rhop     {{\ensuremath{\rhomeson^+}}\xspace}

\def\kaon    {{\ensuremath{\PK}}\xspace}
\def\KorKbar {\kern \thebaroffset\optbar{\kern -\thebaroffset \PK}{}\xspace}

\def\Kp      {{\ensuremath{\kaon^+}}\xspace}
\def\Km      {{\ensuremath{\kaon^-}}\xspace}

\def\Kstarz  {{\ensuremath{\kaon^{*0}}}\xspace}

\newcommand{\phiz}{\ensuremath{\Pphi}\xspace}

\def\Dbar    {{\ensuremath{\offsetoverline{\PD}}}\xspace}
\def\D       {{\ensuremath{\PD}}\xspace}

\def\DorDbar {\kern \thebaroffset\optbar{\kern -\thebaroffset \PD}\xspace}
\def\Dz      {{\ensuremath{\D^0}}\xspace}

\def\Dp      {{\ensuremath{\D^+}}\xspace}
\def\Dm      {{\ensuremath{\D^-}}\xspace}

\def\DpDm    {\ensuremath{\Dp {\kern -0.16em \Dm}}\xspace}
\def\Dstar   {{\ensuremath{\D^*}}\xspace}

\def\Dstarz  {{\ensuremath{\D^{*0}}}\xspace}
\def\Dstarzb {{\ensuremath{\Dbar{}^{*0}}}\xspace}

\def\Dstarp  {{\ensuremath{\D^{*+}}}\xspace}
\def\Dstarm  {{\ensuremath{\D^{*-}}}\xspace}

\def\Ds      {{\ensuremath{\D^+_\squark}}\xspace}

\def\Dsm     {{\ensuremath{\D^-_\squark}}\xspace}

\def\Dss     {{\ensuremath{\D^{*+}_\squark}}\xspace}
\def\Dssp    {{\ensuremath{\D^{*+}_\squark}}\xspace}
\def\Dssm    {{\ensuremath{\D^{*-}_\squark}}\xspace}

\def\B       {{\ensuremath{\PB}}\xspace}
\def\Bbar    {{\ensuremath{\offsetoverline{\PB}}}\xspace}

\def\BorBbar {\kern \thebaroffset\optbar{\kern -\thebaroffset \PB}\xspace}
\def\Bz      {{\ensuremath{\B^0}}\xspace}

\def\Bd      {{\ensuremath{\B^0}}\xspace}

\def\BdorBdbar {\kern \thebaroffset\optbar{\kern -\thebaroffset \Bd}\xspace}
\def\Bu      {{\ensuremath{\B^+}}\xspace}
\def\Bub     {{\ensuremath{\B^-}}\xspace}
\def\Bp      {{\ensuremath{\Bu}}\xspace}
\def\Bm      {{\ensuremath{\Bub}}\xspace}

\def\Bs      {{\ensuremath{\B^0_\squark}}\xspace}
\def\Bsb     {{\ensuremath{\Bbar{}^0_\squark}}\xspace}
\def\BsorBsbar {\kern \thebaroffset\optbar{\kern -\thebaroffset \Bs}\xspace}

\def\jpsi     {{\ensuremath{{\PJ\mskip -3mu/\mskip -2mu\Ppsi}}}\xspace}

\def\Y#1S{\ensuremath{\PUpsilon{(#1S)}}\xspace}

\def\proton      {{\ensuremath{\Pp}}\xspace}
\def\antiproton  {{\ensuremath{\overline \proton}}\xspace}

\def\Lz          {{\ensuremath{\PLambda}}\xspace}
\def\Lbar        {{\ensuremath{\offsetoverline{\PLambda}}}\xspace}
\def\LorLbar     {\kern \thebaroffset\optbar{\kern -\thebaroffset \PLambda}\xspace}

\def\Lc          {{\ensuremath{\Lz^+_\cquark}}\xspace}
\def\Lcbar       {{\ensuremath{\Lbar{}^-_\cquark}}\xspace}

\def\Lb           {{\ensuremath{\Lz^0_\bquark}}\xspace}

\def\BF         {{\ensuremath{\mathcal{B}}}\xspace}
\def\BR         {\BF}

\newcommand{\decay}[2]{\ensuremath{#1\!\to #2}\xspace} 

\def\to                 {\ensuremath{\rightarrow}\xspace}

\newcommand{\mBs}{{\ensuremath{m_{\Bs}}}\xspace}

\def\qsq       {{\ensuremath{q^2}}\xspace}

\def\CP                {{\ensuremath{C\!P}}\xspace}

\def\Vcb  {{\ensuremath{V_{\cquark\bquark}}}\xspace}

\def\AT#1     {\ensuremath{A_{\mathrm{T}}^{#1}}\xspace}           % 2

\def\C#1      {\ensuremath{\mathcal{C}_{#1}}\xspace}                       % 9
\def\Cp#1     {\ensuremath{\mathcal{C}_{#1}^{'}}\xspace}                    % 7
\def\Ceff#1   {\ensuremath{\mathcal{C}_{#1}^{\mathrm{(eff)}}}\xspace}        % 9  
\def\Cpeff#1  {\ensuremath{\mathcal{C}_{#1}^{'\mathrm{(eff)}}}\xspace}       % 7
\def\Ope#1    {\ensuremath{\mathcal{O}_{#1}}\xspace}                       % 2
\def\Opep#1   {\ensuremath{\mathcal{O}_{#1}^{'}}\xspace}                    % 7

\newcommand{\nospaceunit}[1]{\ensuremath{\text{#1}}}       
\newcommand{\aunit}[1]{\ensuremath{\text{\,#1}}}       
\newcommand{\tev}{\aunit{Te\kern -0.1em V}\xspace}
\newcommand{\gev}{\aunit{Ge\kern -0.1em V}\xspace}
\newcommand{\mev}{\aunit{Me\kern -0.1em V}\xspace}
\newcommand{\kev}{\aunit{ke\kern -0.1em V}\xspace}
\newcommand{\ev}{\aunit{e\kern -0.1em V}\xspace}
 
\newcommand{\mevc}{\ensuremath{\aunit{Me\kern -0.1em V\!/}c}\xspace}
\newcommand{\gevc}{\ensuremath{\aunit{Ge\kern -0.1em V\!/}c}\xspace}
\newcommand{\mevcc}{\ensuremath{\aunit{Me\kern -0.1em V\!/}c^2}\xspace}
\newcommand{\gevcc}{\ensuremath{\aunit{Ge\kern -0.1em V\!/}c^2}\xspace}
\def\mum  {\ensuremath{\,\upmu\nospaceunit{m}}\xspace}

\def\pb {\aunit{pb}\xspace}
\def\invpb {\ensuremath{\pb^{-1}}\xspace}
\def\fb   {\ensuremath{\aunit{fb}}\xspace}
\def\invfb   {\ensuremath{\fb^{-1}}\xspace}

\newcommand{\stat}{\aunit{(stat)}\xspace}
\newcommand{\syst}{\aunit{(syst)}\xspace}

\newcommand{\chisq}{\ensuremath{\chi^2}\xspace}

\def\deriv {\ensuremath{\mathrm{d}}}

\def\gsim{{~\raise.15em\hbox{$>$}\kern-.85em
          \lower.35em\hbox{$\sim$}~}\xspace}
\def\lsim{{~\raise.15em\hbox{$<$}\kern-.85em
          \lower.35em\hbox{$\sim$}~}\xspace}

\def\sPlot{\mbox{\em sPlot}\xspace}

\def\pt         {\ensuremath{p_{\mathrm{T}}}\xspace}

\def\ptot       {\ensuremath{p}\xspace}

\def\evtgen     {\mbox{\textsc{EvtGen}}\xspace}

\def\geant      {\mbox{\textsc{Geant4}}\xspace}

\def\photos     {\mbox{\textsc{Photos}}\xspace}

\def\pythia     {\mbox{\textsc{Pythia}}\xspace}

\def\tell1  {TELL1\xspace}
\def\ukl1   {UKL1\xspace}

%% file: title-LHCb-PAPER.tex
% $Id: title-LHCb-PAPER.tex 122889 2018-08-17 17:59:55Z pkoppenb $
% ===============================================================================
% Purpose: LHCb-PAPER journal paper title page template
% Author: 
% Created on: 2010-09-25
% ===============================================================================

%%%%%%%%%%%%%%%%%%%%%%%%%
%%%%%  TITLE PAGE  %%%%%%
%%%%%%%%%%%%%%%%%%%%%%%%%
\begin{titlepage}
\pagenumbering{roman}

% Header ---------------------------------------------------
\vspace*{-1.5cm}
\centerline{\large EUROPEAN ORGANIZATION FOR NUCLEAR RESEARCH (CERN)}
\vspace*{1.5cm}
\noindent
\begin{tabular*}{\linewidth}{lc@{\extracolsep{\fill}}r@{\extracolsep{0pt}}}
\ifthenelse{\boolean{pdflatex}}% Logo format choice
{\vspace*{-1.5cm}\mbox{\!\!\!\includegraphics[width=.14\textwidth]{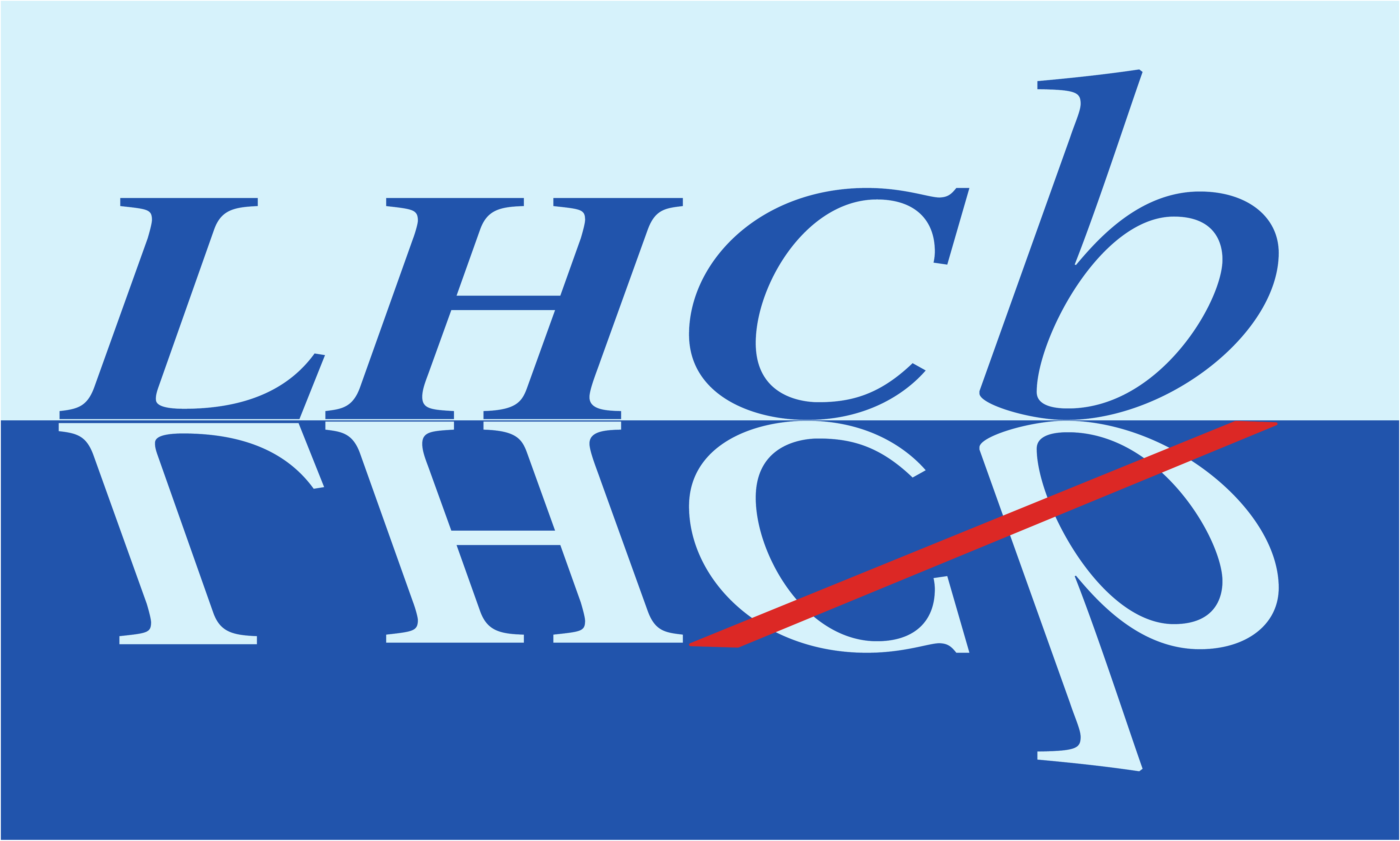}} & &}%
{\vspace*{-1.2cm}\mbox{\!\!\!\includegraphics[width=.12\textwidth]{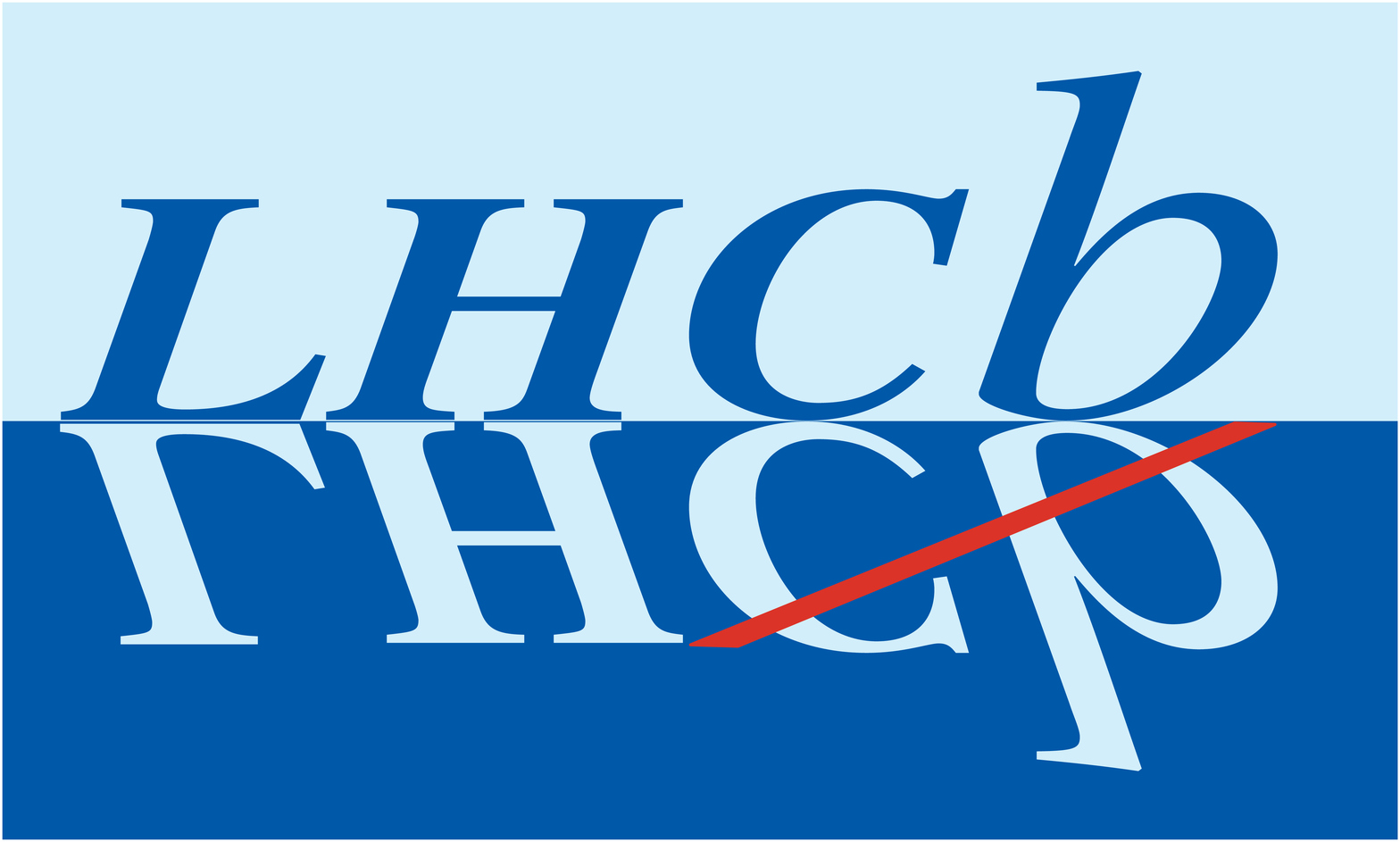}} & &}%
\\
 & & CERN-EP-2020-026 \\  % ID 
 & & LHCb-PAPER-2019-046 \\  % ID 
 & & \today \\ % Date - Can also hardwire e.g.: 23 March 2010
 %& & v3.2 \\
% not in paper \hline
\end{tabular*}

\vspace*{4.0cm}

% Title --------------------------------------------------
{\normalfont\bfseries\boldmath\huge
\begin{center}
% DO NOT EDIT HERE. Instead edit macro in main.tex to keep metadata correct
  \papertitle 
\end{center}
}

\vspace*{2.0cm}

% Authors -------------------------------------------------
\begin{center}
%In the footnote, replace 'paper' by 'Letter' in case of submission to PRL or PLB 
% Edit macro in main.tex to keep metadata correct
\paperauthors\footnote{Authors are listed at the end of this paper.}
\end{center}

\vspace{\fill}

% Abstract -----------------------------------------------
\begin{abstract}
\noindent
The shape of the \mbox{$B_s^0\rightarrow D_s^{*-}\mu^+\nu_{\mu}$} differential decay rate is obtained as a function of the hadron recoil parameter using proton-proton collision data at a centre-of-mass energy of 13\tev, corresponding to an integrated luminosity of 1.7\invfb collected by the \lhcb detector. 
The \mbox{$B_s^0\rightarrow D_s^{*-}\mu^+\nu_{\mu}$} decay is reconstructed through the decays \mbox{$D_s^{*-}\rightarrow D_s^{-}\gamma$} and \mbox{$D_s^{-}\rightarrow K^-K^+\pi^-$}.
The differential decay rate is fitted with the Caprini-Lellouch-Neubert (CLN) and Boyd-Grinstein-Lebed (BGL) parametrisations of the form factors, and the relevant quantities for both are extracted.
%Both the folded and unfolded spectra as a function of the hadronic recoil parameter $w$ are presented. Using both the CLN and BGL parametrisations, a fit to the leading form factor is performed. 

\end{abstract}

\vspace*{2.0cm}

\begin{center}
  Published in JHEP \textbf{12} (2020) 144
\end{center}

\vspace{\fill}

{\footnotesize 
% Edit macro in main.tex to keep metadata correct
\centerline{\copyright~\papercopyright. \href{\paperlicenceurl}{\paperlicence}.}}
\vspace*{2mm}

\end{titlepage}

%%%%%%%%%%%%%%%%%%%%%%%%%%%%%%%%
%%%%%  EOD OF TITLE PAGE  %%%%%%
%%%%%%%%%%%%%%%%%%%%%%%%%%%%%%%%

%  empty page follows the title page ----
\newpage
\setcounter{page}{2}
\mbox{~}
%\newpage
%
%% Author List ----------------------------
%%  You need to get a new author list!
%\input{LHCb_Authorship_17-Dec-2019.tex}
%
%The author list for journal publications is provided by the Membership Committee shortly after 'approval to go to paper' has been given.
%%It will be made available on the page
%%\verb!http://www.physik.uzh.ch/~strauman/forMemCo/LHCb-PAPER-XXXX-XXX/! .
%It will be sent to you by email shortly after a paper number has beens assigned.
%The author list should be included already at first circulation, 
%to allow new members of the collaboration to verify whether they have been included correctly.
%Occasionally a misspelled name is corrected or associated institutions become full members.
%In that case, a new author list will be sent to you.
%In case line numbering doesn't work well after including the authorlist, try moving the \verb!\bigskip! after the last author to a separate line.
%
%
%The authorship for Conference Reports should be ``The LHCb
%  collaboration'', with a footnote giving the name(s) of the contact
%  author(s), but without the full list of collaboration names.

\cleardoublepage

%% file: introduction.tex
\section{Introduction}
\label{sec:Introduction}
Semileptonic decays of heavy hadrons are commonly used to measure the parameters of the Cabibbo-Kobayashi-Maskawa (CKM) matrix~\cite{Cabibbo:1963yz,Kobayashi:1973fv}, 
as they involve only one hadronic current that can be parametrised in terms of scalar functions known as form factors. The number of form factors needed to describe a particular decay depends upon the spin of the initial- and final-state hadrons~\cite{Fakirov:1977ta,Bauer:1986bm,Wirbel:1985ji}. 
For the decay of a pseudoscalar \B meson to a vector \Dstar meson, four form factors are required. 
The determination of the CKM matrix element $|V_{cb}|$ using \mbox{$B\rightarrow \D^{(*)}\ell\neul$} decays or via the inclusive sum of all hadronic $B\to X_c\ell\nu_\ell$ decay channels has been giving inconsistent results during the last thirty years~\cite{HFLAV18}. The exclusive determination relies heavily on the parametrisation of the form factors, as it requires an extrapolation of the differential decay rate to the zero recoil point, where the momentum transfer to the lepton system is maximum.

Recently, the \lhcb collaboration has measured $|V_{cb}|$ using  $\Bs\rightarrow \D_s^{(*)-}\mup\nu_\mu$ decays\footnote{The inclusion of charge-conjugate processes is implied throughout this paper.} with two form-factor parametrisations, giving consistent results~\cite{LHCb-PAPER-2019-041}. The determination of the form factors in $\Bs\to\Dssm\ellp\neul$ decays obtained using different parametrisations can help to clarify the $|V_{cb}|$ inconsistency between the exclusive and inclusive approaches. It can also be used to improve the Standard Model (SM) predictions of the \BsToDsstaunu branching fraction and the ratio $\mathcal{R}(D_s^*)=\BR(\BsToDsstaunu)/\BR(\BsToDssmunu)$. A measurement and precise prediction of the latter could increase the understanding of the current tension between experimental and theoretical values of the equivalent ratio $\mathcal{R}(\D^{(*)})=\BR(\B\rightarrow \D^{(*)}\taup\neut)/\BR(\B\rightarrow \D^{(*)}\mup\neum)$~\cite{HFLAV18}. Theoretical predictions on \Bs semileptonic decays are expected to be more precise than those on \Bz or \Bp decays. For example, the Lattice QCD calculations of the form factors are computationally easier due to the larger mass of the spectator \squark quark compared to that of \uquark or \dquark quarks \cite{Monahan:2017uby, Harrison:2017fmw}.
Despite these advantages, the study of semileptonic \Bs decays has received less theoretical attention than the equivalent \Bd and \Bu decays
due to the lack of experimental results. 

This paper reports the first measurement of the shape of the differential decay rate of the \BsToDssmunu decay as a function of the hadronic recoil parameter
\mbox{$w = v_{\Bs}\cdot v_{\Dssm}$}, 
where $v_{\Bs}$ and $v_{\Dssm}$ are the four-vector velocities of the \Bs and \Dssm mesons, respectively.
The spectrum of $w$ is unfolded accounting for the detector resolution on $w$ and corrected for the reconstruction and selection efficiency. The \Dssm meson is reconstructed in the \mbox{$D_s^{*-}\rightarrow D_s^{-}\gamma$} mode, where the $D_s^{-}$ meson subsequently decays via the
\mbox{$D_s^{-}\rightarrow \phi(\to \Kp\Km)\pi^-$}
or \mbox{$D_s^{-}\rightarrow \Kstarz (\to \pim \Kp) \Km$} mode. The data used correspond to an integrated luminosity of 1.7\invfb collected by the LHCb experiment in 2016 at a centre-of-mass energy of 13\tev.
 
The \BsToDssmunu decay is described by four form factors.
The most commonly used parametrisations to model these form factors are by Caprini-Lellouch-Neubert (CLN)~\cite{Caprini:1997mu} and by Boyd-Grinstein-Lebed (BGL)~\cite{Boyd:1994tt,Boyd:1995cf,Boyd:1995sq}. 
This paper also describes how the relevant parameters of these parametrisations can be extracted by fitting the measured differential decay rate.

\section{Formalism of the \texorpdfstring{\boldmath\BsToDssmunu}{Bs->Ds*munu} decay}
\label{sec:eqs}

\begin{figure}[tb]
  \centering
  \includegraphics[width=0.6\textwidth]{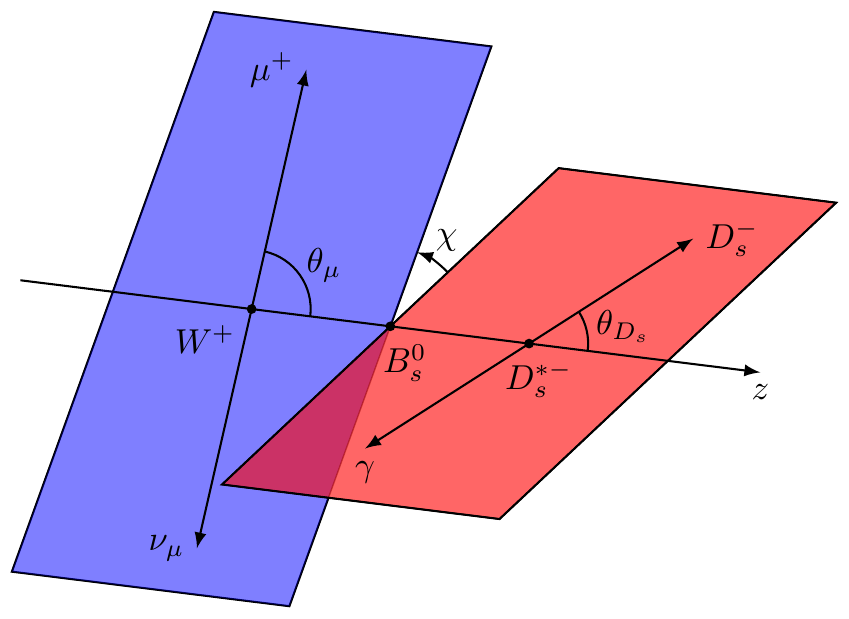}
  \caption{Schematic overview of the \BsToDssmunu decay, introducing the angles \thetaV, \thetal \mbox{and $\chi$}.}
  \label{fig:angles}
\end{figure}

The \BsToDssmunu decay, with the subsequent \DssToDsg decay, can be described by three angular variables and the squared momentum transfer to the lepton system, defined as $\qsq=(\ptot_{\Bs}-\ptot_{\Dssm})^2$, where $\ptot_{\Bs}$ and $\ptot_{\Dssm}$ are the four-momenta of the \Bs and \Dssm mesons, respectively. The three angular variables, indicated in \Figref{fig:angles}, are two helicity angles \thetal and \thetaV, and the angle $\chi$. In this figure the direction of the $z$-axis is defined in the \Bs rest frame as $\hat{z} =  \vec{p}_{\Dssm}/|\vec{p}_{\Dssm}|$. The angle between the muon direction in the virtual \W rest frame and the $z$ direction is called \thetal, while the angle between the \Dsm meson direction in the \Dssm rest frame and the $z$ direction is called \thetaV. Finally, $\chi$ is the angle between the plane formed by the 
\Dssm decay products and that formed by the two leptons in the \Bs rest frame~\cite{Colangelo:2018cnj}. 
The angular basis is designed such that the angular definition for the \Bsb decay is a \CP transformation of that of the \Bs decay.

The measurement is performed by integrating the full decay rate over the decay angles. Thus, the expression of the \mbox{\BsToDssmunu} decay rate is given by
\begin{equation}
\begin{split}
\frac{\deriv\Gamma(\BsToDssmunu)}{\deriv\qsq} &= \frac{G^2_{\rm F} \, |\Vcb|^2 \, |\eta_{\rm EW}|^2 \, |\vec{p}\,| \qsq} { 96 \, \pi^3 \, \mBssq}
    \left( 1- \frac{m_{\mu}^2}{\qsq} \right)^2 \\
    &\quad \times \left[ (|H_+|^2 + |H_-|^2 + |H_0|^2) \left( 1+ \frac{m_{\mu}^2}{2 \, \qsq} \right) 
    +\frac{3}{2} \frac{m_{\mu}^2}{\qsq} |H_{t}|^2 \right] \, .
\end{split}
\label{eq:decayrate}
\end{equation}
In this equation, $G_{\rm F}$ is the Fermi constant, \Vcb is the CKM matrix element describing the \bquark to \cquark transition, \mbox{$\eta_{\rm EW} = 1.0066$} is the  
electroweak correction to \Vcb~\cite{Sirlin:1981ie}, $m_{\mu}$ is the muon mass~\cite{PDG2019},
and $H_0$, $H_+$, $H_-$, $H_{t}$ are the helicity amplitudes.
The magnitude of the \Dssm momentum in the \Bs rest frame is given by $|\vec{p}|$. The hadronic recoil, $w$, is related to the squared momentum transfer to the lepton pair, \qsq, by
\begin{equation}
w = \frac{\ptot_{\Bs}}{m_{\Bs}}\cdot\frac{\ptot_{\Dssm}}{m_{\Dssm}} = \frac{m^2_{\Bs}+m^2_{\Dssm}-\qsq}{2\;m_{\Bs}\;m_{\Dssm}},
\label{eq:w}
\end{equation}
where $m_{\Bs}$ and $m_{\Dssm}$ are the masses of the \Bs and \Dssm mesons, respectively.
 The minimal value, $\w=1$, corresponds to the situation in which the \Dssm meson has zero recoil in the \Bs rest frame. It is also the value for which \qsq is maximal.

The dependence of the helicity amplitudes on $w$ can be
expressed in different ways, most commonly described in either the CLN or BGL parametrisations, as discussed further in \Secref{sec:CLN} and \Secref{sec:BGL}. This analysis is only sensitive to a single form-factor contribution while the other form factors are fixed to existing measurements from \Bp and \Bd semileptonic decays~\cite{HFLAV18,Gambino:2019sif}.
This is supported by Ref.~\cite{Kobach:2019kfb}, where when imposing unitarity and analyticity the differences in form factors for semileptonic $B\rightarrow D$ and $\Bs\rightarrow \Ds$ decays are found to be within $\mathcal{O}(1\%)$ over the entire kinematic range. Also, a simultaneous analysis of the $B_q\rightarrow D_q^{(*)}$ form factors for both light ($q=u,d$) and strange ($q=s$) spectator quarks within the Heavy-Quark-Expansion framework to order $\mathcal{O}(\alpha_s,1/m_b,1/m_c^2)$~\cite{Bordone:2019guc} does not show any significant SU(3) symmetry breaking.
Moreover, Lattice QCD calculations indicate that there is also good agreement of the form factors at zero recoil~\cite{Harrison:2017fmw,McLean:2019sds}.

\subsection{CLN form-factor parametrisation}
\label{sec:CLN}
For the CLN parametrisation~\cite{Caprini:1997mu}, the helicity amplitudes $H_0$, $H_+$, $H_-$ and $H_{t}$ can be written in terms of the form factors 
$A_1(w)$, $V(w)$, $A_2(w)$ and $A_0(w)$ as
\begin{align}
H_{\pm}(w) &= \mBs \, (1+r) \, A_1(w) \mp \frac{2}{1 + r} \, |\vec{p}\,| \, V(w) , \notag
\\[10pt]
H_0(w) &= \frac{\mBs \, \mDss \, (w-r)\,  (1+r)^2\,A_1(w)-2\,|\vec{p}\,|^2\, A_2(w)}{\mDss \, (1+r)\, \sqrt{1+r^2-2wr}} \, , 
\\[10pt]
H_{t}(w) &= \frac{2 \, |\vec{p}\,|}{\sqrt{1+r^2-2wr}} \, A_0 (w) \, , \notag
\label{eq:H}
\end{align}
where $r=\mDss/\mBs$.
The form factors are rewritten in terms of a single leading form factor 
\begin{equation}
h_{A_1} (w) = A_1(w) \frac{1}{R_{\Dssm}}\frac{2}{w+1} \, ,\\
\label{eq:define_hA}
\end{equation}
and three ratios of form factors
\begin{equation}
R_0 (w)=\frac{A_0(w)}{h_{A_1}(w)}R_{\Dssm} \, , \qquad
R_1 (w)=\frac{V(w)}{h_{A_1}(w)}R_{\Dssm} \, , \qquad
R_2 (w)=\frac{A_2(w)}{h_{A_1}(w)}R_{\Dssm} \, , 
\label{eq:define_Ri}
\end{equation}
where 
\begin{equation}
R_{\Dssm} = \frac{2 \sqrt{r}}{1+r}.
\end{equation}

In the CLN parametrisation, the leading form factor and the three ratios
are parametrised in terms of \w as
\begin{equation}
\begin{aligned}
h_{A_{1}}(w) &= h_{A_{1}}(1)[1 - 8\rho^{2}z(w) +(53\rho^{2} - 15)z^2(w) - (231\rho^{2} - 91)z^3(w)] \ , \\
R_{0}(w) &= R_{0}(1) - 0.11(w - 1) + 0.01(w - 1)^{2} \ , \\
R_{1}(w) &= R_{1}(1) - 0.12(w - 1) + 0.05(w - 1)^{2} \ , \\
R_{2}(w) &= R_{2}(1) + 0.11(w - 1) - 0.06(w - 1)^{2} \ , 
\label{eq:CLN:param}
\end{aligned}
\end{equation}
where the coefficients, originally calculated for \B decays, are assumed to be the same for \Bs decays.
The function $z(w)$ is defined as
\begin{equation}
z(w) = \frac{\sqrt{w +1} - \sqrt{2}}{\sqrt{w + 1} + \sqrt{2}} .
\end{equation}
As this analysis is only sensitive to the shape of the form-factor parametrisation the term $h_{A_1}(1)$ is absorbed in the normalisation.
 The values of $R_{1}(1)$ and $R_{2}(1)$ are taken from the HFLAV average of the corresponding  parameters, obtained from $B\to D^*\ell\nu_\ell$ decays~\cite{HFLAV18}.
The $R_{0}(1)$ parameter is suppressed by $m_{\ell}^{2}/q^2$ in the helicity amplitude and its contribution to the total rate is negligible. The value predicted by the exact heavy quark limit of $R_{0}(1)=1$~\cite{Fajfer:2012vx} is used, as no measurement of $R_0(1)$ has been performed.
The slope, $\rho^2$, of $h_{A_1}(w)$ is the only parameter fitted in this parametrisation.

\subsection{BGL form-factor parametrisation}
\label{sec:BGL}
In the BGL parametrisation~\cite{Boyd:1994tt,Boyd:1995cf,Boyd:1995sq}, the helicity amplitudes are parametrised as
\begin{align}
H_0(w) &= \frac{\mathcal{F}_{1}(w)}{\mBs \, \sqrt{1+r^2+2wr}} \ , \notag
\\[10pt]
H_{\pm}(w) &=  f(w) \mp \mBs\,\mDss\sqrt{w^2 -1}g(w) \ , 
\\[5pt]
H_{t}(w) &=  m_{\Bs} \frac{\sqrt{r}(1 + r)\sqrt{w^2 -1}}{\sqrt{1 + r^2 - 2 w r}}\mathcal{F}_{2}(w)  \notag \ ,
\label{eq:Hel_amp_BGL}
\end{align}
where the form factors are defined as
\begin{equation}
\begin{aligned}
f(z) &= \frac{1}{P_{1^{+}}(z)\phi_{f}(z)}\sum_{n=0}^{\infty}a_{n}^{f}z^{n} \ , \qquad &
\mathcal{F}_{1}(z) &= \frac{1}{P_{1^{+}}(z)\phi_{\mathcal{F}_{1}}(z)}\sum_{n=0}^{\infty}a_{n}^{\mathcal{F}_{1}}z^{n} \ , \\ 
g(z) &= \frac{1}{P_{1^{-}}(z)\phi_{g}(z)}\sum_{n=0}^{\infty}a_{n}^{g}z^{n} \ , &
\mathcal{F}_{2}(z) &= \frac{\sqrt{r}}{(1+r)P_{0^{-}}(z)\phi_{\mathcal{F}_{2}}(z)}\sum_{n=0}^{\infty}a_{n}^{\mathcal{F}_{2}}z^{n} \ . 
\label{eq:Form_factors_BGL}
\end{aligned}
\end{equation}
The functions $\phi_i$ are the so-called outer functions,  $P_{1^{\pm}, 0^-}$ are Blaschke factors, and the coefficients $a_n^i$, where $i=\{f,\,g,\,\mathcal{F}_1,\,\mathcal{F}_2\}$, are parameters to be fit from data.

As the form-factor parametrisation is given through analytic functions, they must satisfy the unitarity condition in the $z$ expansion
\begin{equation}
    \sum_{n=0}^{\infty} (a_n^g)^2 \leq 1 \, , \qquad
    \sum_{n=0}^{\infty} (a_n^f)^2 + \sum_{n=0}^{\infty} (a_n^{\mathcal{F}_1})^2 \leq 1 \, , \qquad 
    \sum_{n=0}^{\infty} (a_n^{\mathcal{F}_2})^2 \leq 1. \label{eq:unitarity}
\end{equation}
This analysis is only sensitive to the form factor $f(z)$, and its
series is truncated at $n=2$, following Refs.~\cite{Bigi:2017njr,Grinstein:2017nlq,Bigi:2017jbd,Gambino:2019sif,Jaiswal:2020wer}.
The shapes for $\mathcal{F}_1(z)$ and $g(z)$ are constrained using the results in Ref.~\cite{Gambino:2019sif}, where the $a_n^i$ coefficients are fit using recent Belle measurements with $\Bd\to\Dstarm\ell^+\nu_\ell$ decays~\cite{Abdesselam:2017kjf,Abdesselam:2018nnh}. The value of $a_0^f$ in Ref.~\cite{Gambino:2019sif} is determined from the combination of lattice calculations in Ref.~\cite{Aoki:2019cca}. The parameters $a_n^{\mathcal{F}_2}$ for $\mathcal{F}_2(z)$ are fixed from predictions in Ref.~\cite{Bigi:2017jbd}, where they are called $P_1$. As this analysis represents the first measurement of form factors in \Bs\to\Dssm transitions, the choice of the input parameters is driven by having as much experimental input as possible. An overview of the fit inputs is given in \Tabref{tab:BGLinputs} in App.~\ref{sec:inputsBGL}.

%% file: detector.tex
\section{Detector and simulation}
\label{sec:Detector}
The \lhcb detector~\cite{LHCb-DP-2008-001, LHCb-DP-2014-002} is a single-arm forward
spectrometer covering the \mbox{pseudorapidity} range $2<\eta <5$,
designed for the study of particles containing \bquark or \cquark
quarks. The detector includes a high-precision tracking system
consisting of a silicon-strip vertex detector surrounding the $pp$
interaction region~\cite{LHCb-DP-2014-001}, a large-area silicon-strip detector located
upstream of a dipole magnet with a bending power of about
$4{\mathrm{\,Tm}}$, and three stations of silicon-strip detectors and straw
drift tubes~\cite{LHCb-DP-2017-001} placed downstream of the magnet.
The tracking system provides a measurement of the momentum, \ptot, of charged particles with
a relative uncertainty that varies from 0.5\% at low momentum to 1.0\% at 200\gevc.
The minimum distance of a track to a primary vertex (PV), the impact parameter (IP), 
is measured with a resolution of $(15+29/\pt)\mum$,
where \pt is the component of the momentum transverse to the beam, in\,\gevc.
Different types of charged hadrons are distinguished using information
from two ring-imaging Cherenkov detectors~\cite{LHCb-DP-2012-003}. 
Photons, electrons and hadrons are identified by a calorimeter system consisting of
scintillating-pad and preshower detectors, an electromagnetic
and a hadronic calorimeter. Muons are identified by a
system composed of alternating layers of iron and multiwire
proportional chambers~\cite{LHCb-DP-2012-002}.
The online event selection is performed by a trigger~\cite{LHCb-DP-2012-004}, 
which consists of a hardware stage, based on information from the calorimeter and muon
systems, followed by a software stage, which applies a full event
reconstruction. The hardware muon trigger selects events containing a high-\pt muon candidate. The software trigger requires three tracks with a significant displacement from any primary $pp$ interaction vertex.

Simulation is required to model the effects of the detector acceptance and the imposed selection requirements.
In the simulation, $pp$ collisions are generated using \pythia~\cite{Sjostrand:2006za, *Sjostrand:2007gs}  with a specific \lhcb
configuration~\cite{LHCb-PROC-2010-056}.  Decays of unstable particles
are described by \evtgen~\cite{Lange:2001uf}, in which final-state
radiation is generated using \photos~\cite{Golonka:2005pn}. The
interaction of the generated particles with the detector, and its response,
are implemented using the \geant
toolkit~\cite{Allison:2006ve, *Agostinelli:2002hh} as described in
Ref.~\cite{LHCb-PROC-2011-006}.

The simulation is corrected for mismodeling of the kinematic properties of
the generated \Bs mesons and of the photons from the \Dssm decays, 
as well as for data-simulation differences in the muon trigger efficiency and tracking efficiencies of the final-state particles. 
Corrections to the \Bs and \g kinematic distributions are determined by comparing data and simulated samples of \BToJpsiK and \BsToDsstPi decays, respectively. 
Kinematic differences between
\Bs and $B^+$ mesons due to their production mechanisms are small and considered to be negligible~\cite{fsfd,LHCb-PAPER-2019-020}.
Corrections to the trigger and tracking efficiencies are evaluated using data and simulated samples of \BToJpsiK decays~\cite{LHCb-PUB-2014-039}. 
In the simulated signal sample, the form factors are described following the CLN parametrisation with numerical values $\rho^2=1.205$, $R_1(1)=1.404$ and $R_2(1)=0.854$. 

%% file: selection.tex
\section{Data selection}
\label{sec:selection}
Candidate $\Bs\to\Dssm\mup\neum$ decays are selected by pairing \Dssm and \mup candidates, where the \Dssm candidate is reconstructed through the \Dsm\g decay.
The \Dsm mesons are reconstructed requiring two opposite-sign kaons and a pion inconsistent with coming from a PV, and forming a common vertex that is displaced from every PV.
The final-state hadrons and muon must satisfy strict particle identification (PID) criteria, consistent with the assigned particle hypothesis. 

To suppress the combinatorial background in the \Dsm mass spectrum, only the regions of the $\Dsm\to\Kp\Km\pim$ Dalitz plot compatible with originating from the $\phiz\pim$ and $\Kstarz\Km$ decay modes are retained by requiring the \Kp\Km mass to be within $20\mevcc$ of the known \Pphi mass, or the reconstructed \Kp\pim mass to be within $90\mevcc$ of the average $K^*(892)^0$ mass~\cite{PDG2019}. 
Possible backgrounds arising from the misidentification of one of the \Dsm decay products are removed with explicit vetoes which apply more stringent PID requirements in a small window of invariant mass of the corresponding particle combination. The main contributions that are removed come from $\Lcbar \to \Kp \antiproton \pim$, $\Dm \to \Kp \pim \pim$, \mbox{$\Dsm \to \Km \pip \pim$}, and misidentified or partially reconstructed multibody \D decays, all originating from semileptonic \bquark-hadron decays. 

Due to the small mass difference between the \Dssm and \Dsm mesons, the photon must be emitted close to the \Dsm flight direction. Photons are selected inside a narrow cone surrounding the \Dsm candidate, defined in pseudorapidity and azimuthal angle. Only the highest \pt photon inside the cone is combined  with the \Dsm candidate. Potential contamination from neutral pions reconstructed as a single merged cluster in the electromagnetic calorimeter is suppressed by employing a neural network 
classifier trained to separate \piz mesons from photons~\cite{CalvoGomez:2042173}. 

\begin{figure}[t!]
  \centering
  \includegraphics[width=0.8\textwidth]{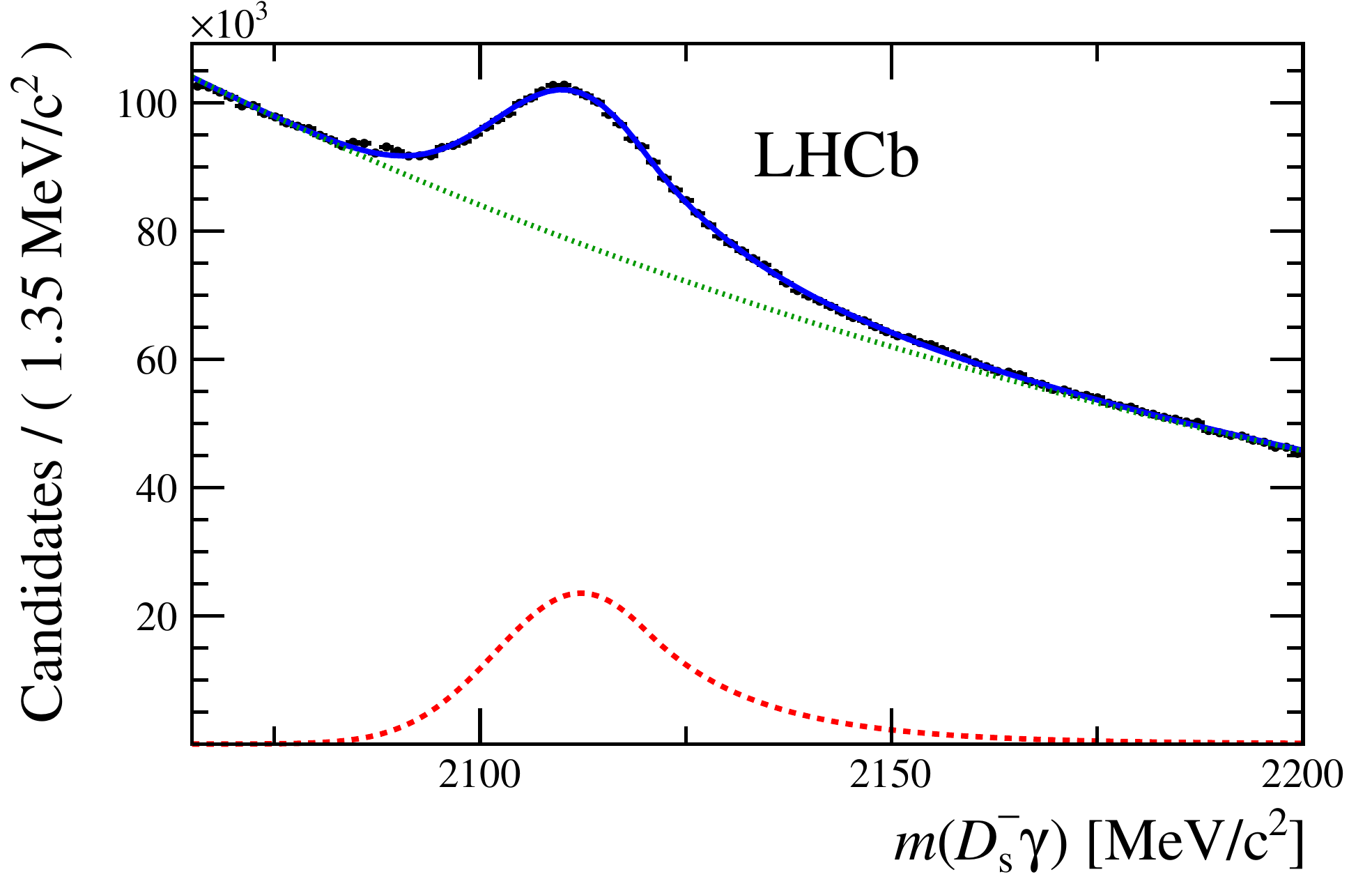}
  \caption{Distribution of the reconstructed $\Dsm\g$ mass, $m(\Dsm\g)$, with the fit overlaid. The fit is performed constraining the \Dsm mass to the world-average value~\cite{PDG2019}. The signal and background components are shown separately with dashed red and dotted green lines, respectively.}
  \label{fig:sWeight}
\end{figure}

A fit to the $\Dsm\g$ invariant-mass distribution, with the reconstructed \Dsm mass constrained to the known value~\cite{PDG2019}, is performed as shown in \Figref{fig:sWeight}. The signal is described by a  Gaussian function with a power-law tail on the right hand side of the distribution and the background by an exponential distribution. The power-law tail accounts for cases where additional activity in the calorimeter is mistakenly included in the photon cluster. The \sPlot technique~\cite{Pivk:2004ty} is employed
to subtract the combinatorial background from random photons. Weighted signal is used to create the templates described in \Secref{sec:yield}. The correlation between the weights and \w is below 4\%. 

The muon candidate is required to have \pt in excess of 1.2\gevc. Background arising from \bquark-hadrons decaying into final states containing two charmed hadrons, \HbToDssHc, followed by a semileptonic decay of the charmed hadron \HcTomunu, where $X$ is one or more hadrons, are suppressed by using a multivariate algorithm based on the isolation of the muon~\cite{LHCB-PAPER-2015-031}.
Finally, the \Bs meson candidates are formed by combining \mup and \Dssm candidates which are consistent with coming from a common vertex.

%% file: yields.tex
\section{Signal yield}
\label{sec:yield}

The signal yield is determined using a template fit to the distribution of the corrected mass~\cite{LHCb-PAPER-2015-013}, 
\begin{equation}
\mcorr = \sqrt{m^2_{\Dssm\mup}+|p_{\perp}|^2} + |p_{\perp}| ,
\end{equation}
where $m_{\Dssm\mup}$ is the measured mass of the \Dssm\mup candidate, and $p_{\perp}$ is the momentum of the candidate transverse to the \Bs flight direction. When only one massless final-state particle is missing from the decay, \mcorr peaks at the \Bs mass. 
Only candidates in the range \mbox{$3500 < \mcorr < 5367\mevcc$} are considered. 

Extended binned maximum-likelihood fits to the \mcorr distribution are performed independently in seven bins of the reconstructed hadronic recoil, $w$, 
to obtain the raw yields $N_{\mathrm{meas}}$ per bin. The binning scheme, detailed in \Tabref{tabs:bins}, is chosen such that each $w$ bin has roughly the same signal yield, based on simulation. Obtaining the value of $w$ requires 
the knowledge of the momentum of the \Bs meson, which in the decays under study can be solved up to a quadratic ambiguity. By imposing momentum balance against the visible system with respect to the flight direction, and assuming the mass of 
the \Bs meson, the momentum of the \Bs meson can be estimated. To resolve the ambiguity 
in the solutions, a multivariate regression algorithm based on the flight direction is used~\cite{Ciezarek:2016lqu} yielding a purity on the solutions of around 70\%. The \mcorr distribution is fitted using shapes (templates) of signal and of background distributions mostly obtained from simulation. These simulated decays are selected as described in~\Secref{sec:selection},
and are corrected for the simulation mismodeling as described in~\Secref{sec:Detector}.

\input{tabs/bins.tex}

The largest contribution to the background is due to \BsToDsstaunu decays, with \mbox{\TauToMu}.
A small source of background is formed by excited \Dsm mesons decaying into a \Dssm resonance. The only such excited state is the \Dsonem meson, and hence templates for \mbox{\BsToDsonemunu}
and \BsToDsonetaunu decays are included in the fit.
The background arising from 
\bquark hadrons decaying into final states containing two charmed hadrons, 
\HbToDssHc, is also addressed.
The template for this 
process is generated using simulated events of \Bs, \Bd, \Bu and \Lb decays, with an 
appropriate admixture of final states, based on their production rates, branching ratios and 
relative reconstruction efficiencies taken from simulation.
The last background considered in the fit is the combinatorial background, arising from random combinations
of tracks. This template is obtained from a data sample where the \Dsm meson and the muon have the same charge.  

\begin{figure}[p]
  \centering
  \includegraphics[width=0.49\textwidth]{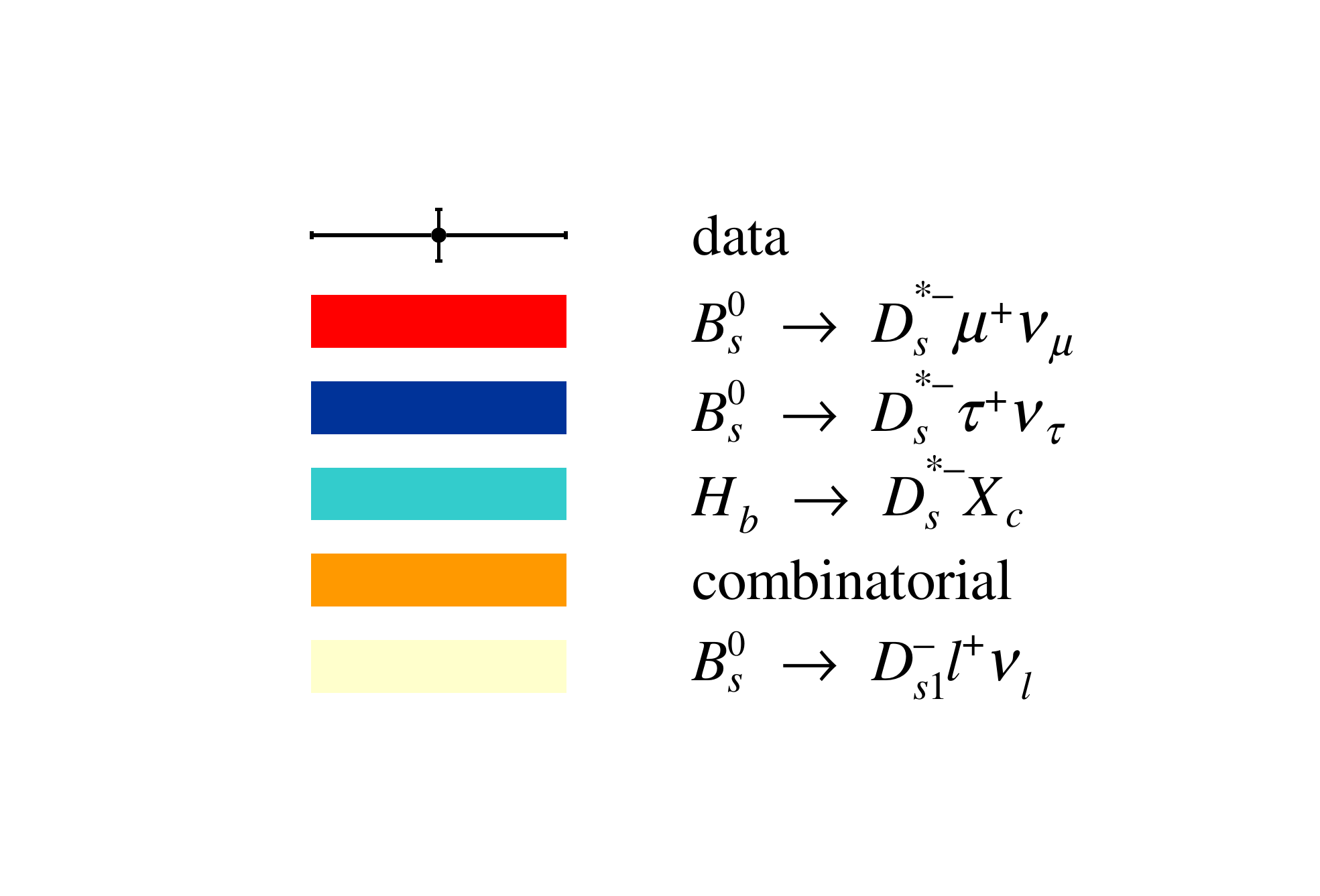}
  \includegraphics[width=0.49\textwidth]{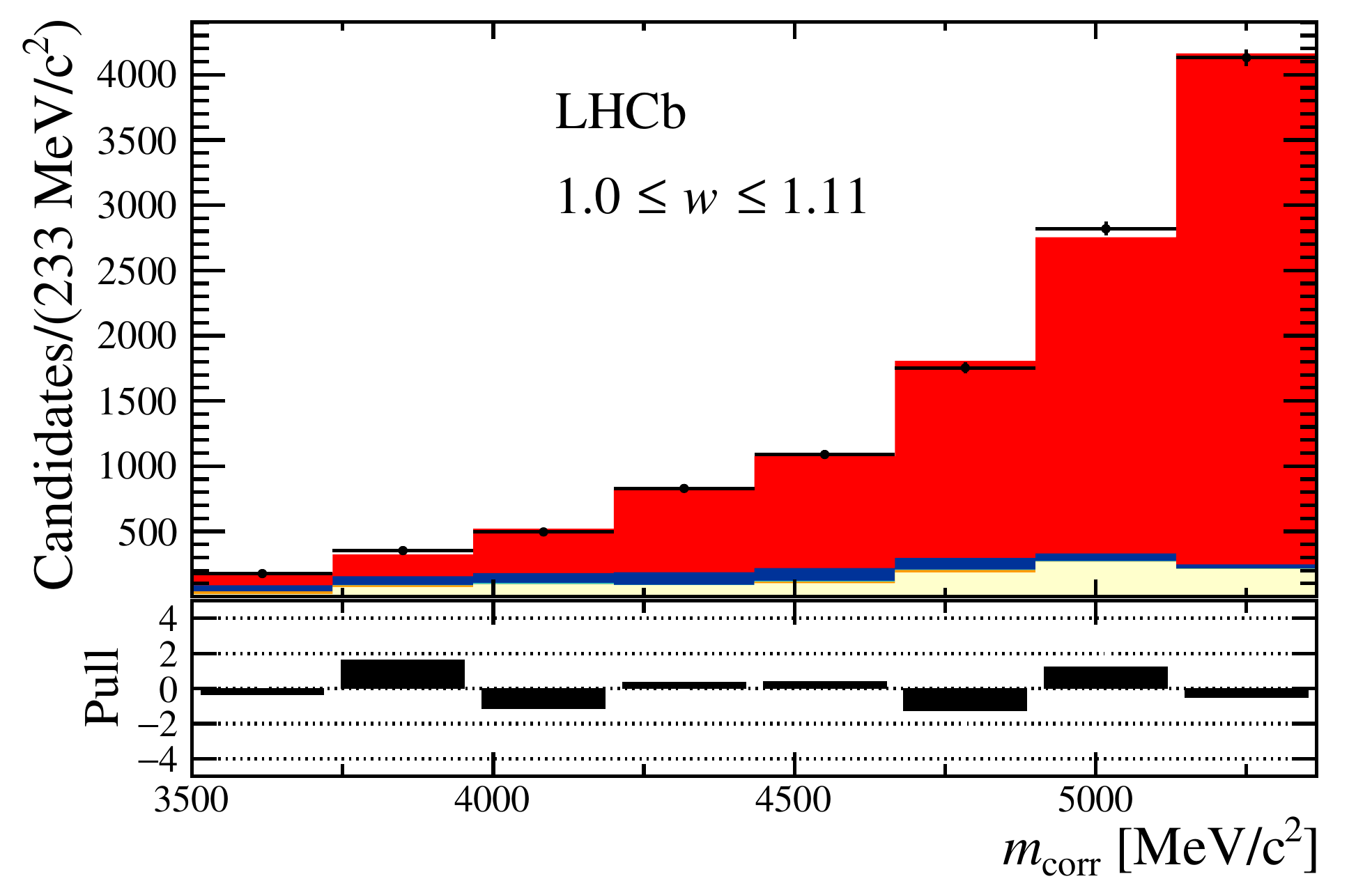}
  \includegraphics[width=0.49\textwidth]{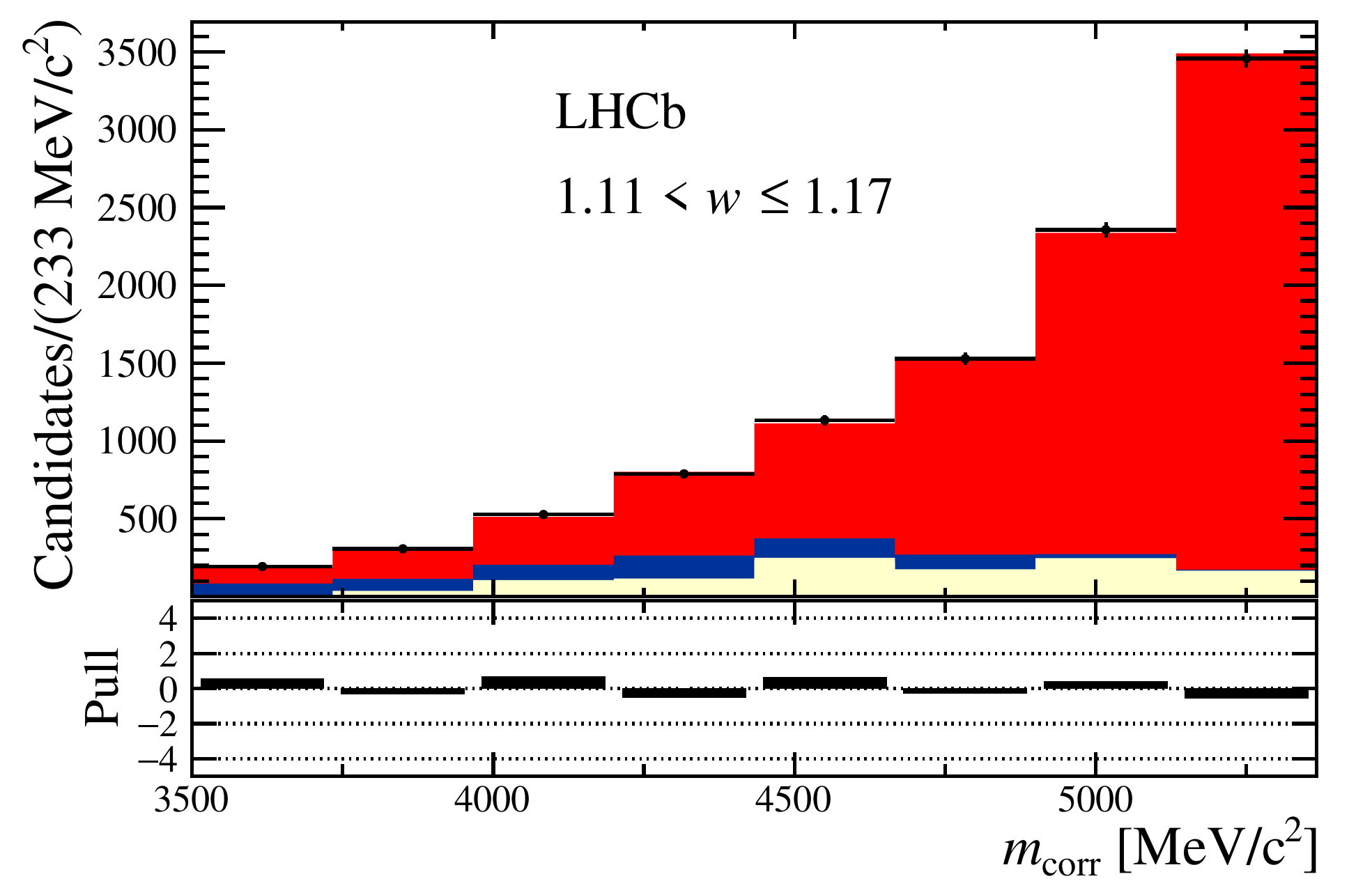}
  \includegraphics[width=0.49\textwidth]{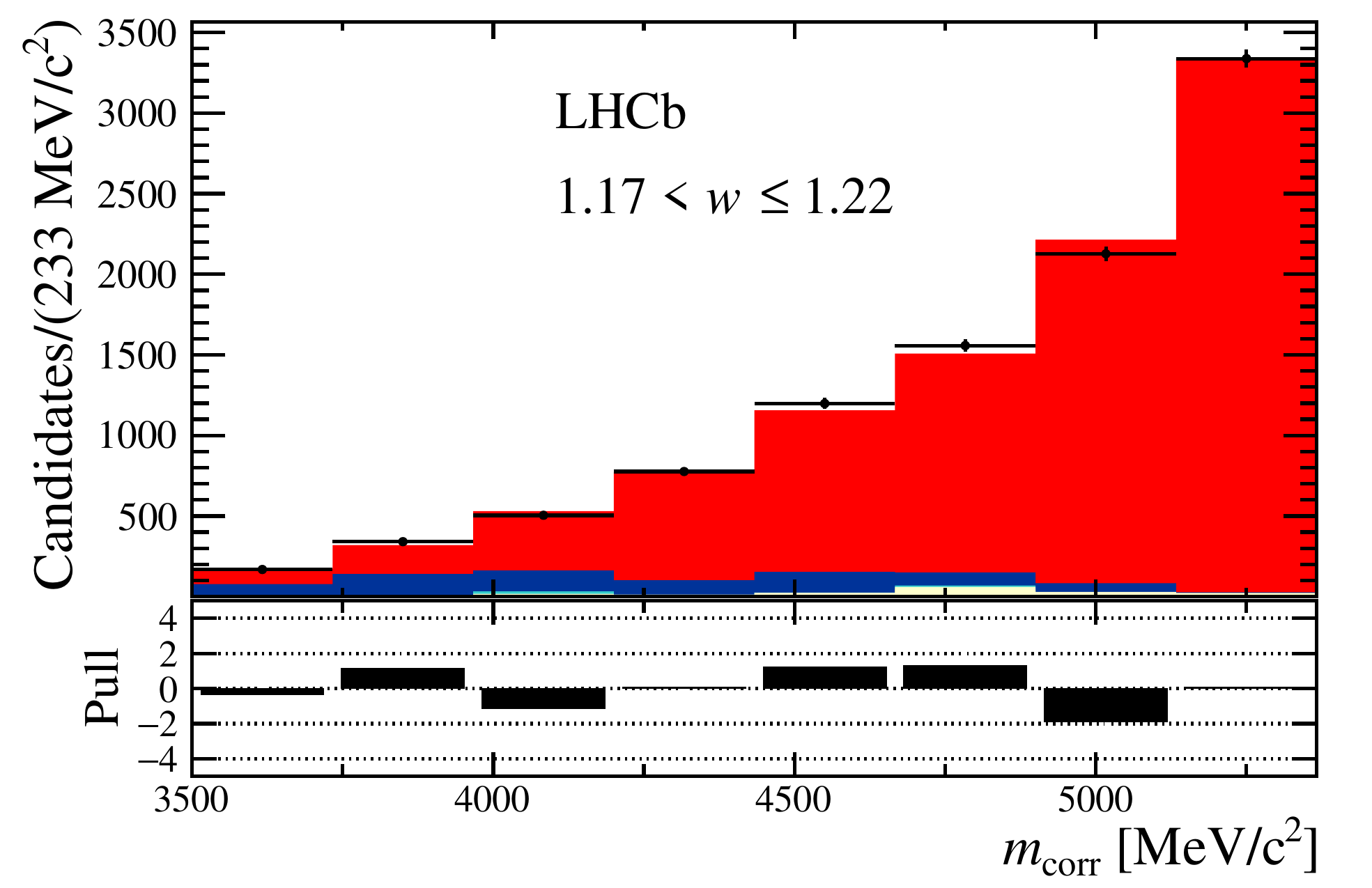}
  \includegraphics[width=0.49\textwidth]{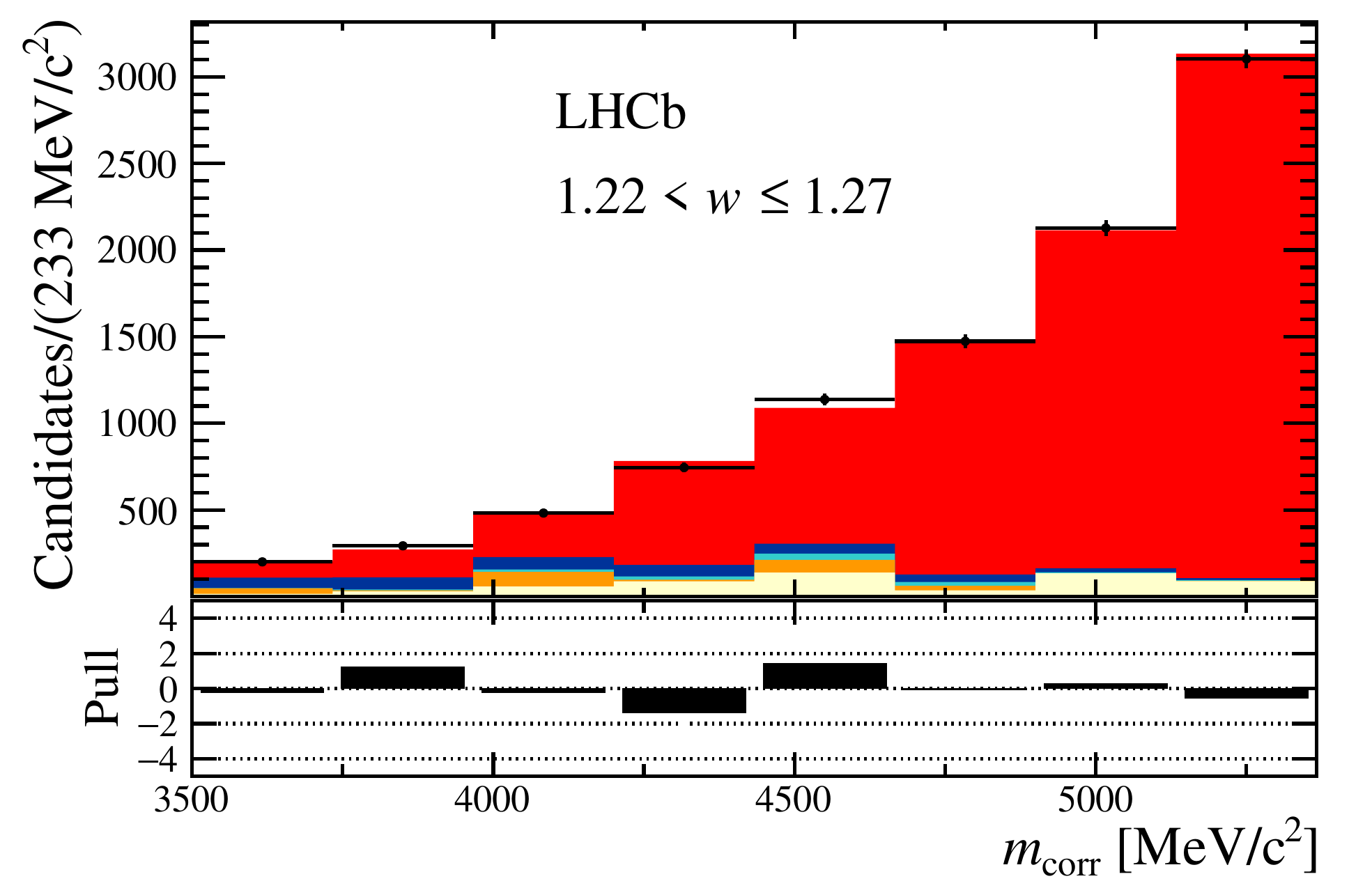}
  \includegraphics[width=0.49\textwidth]{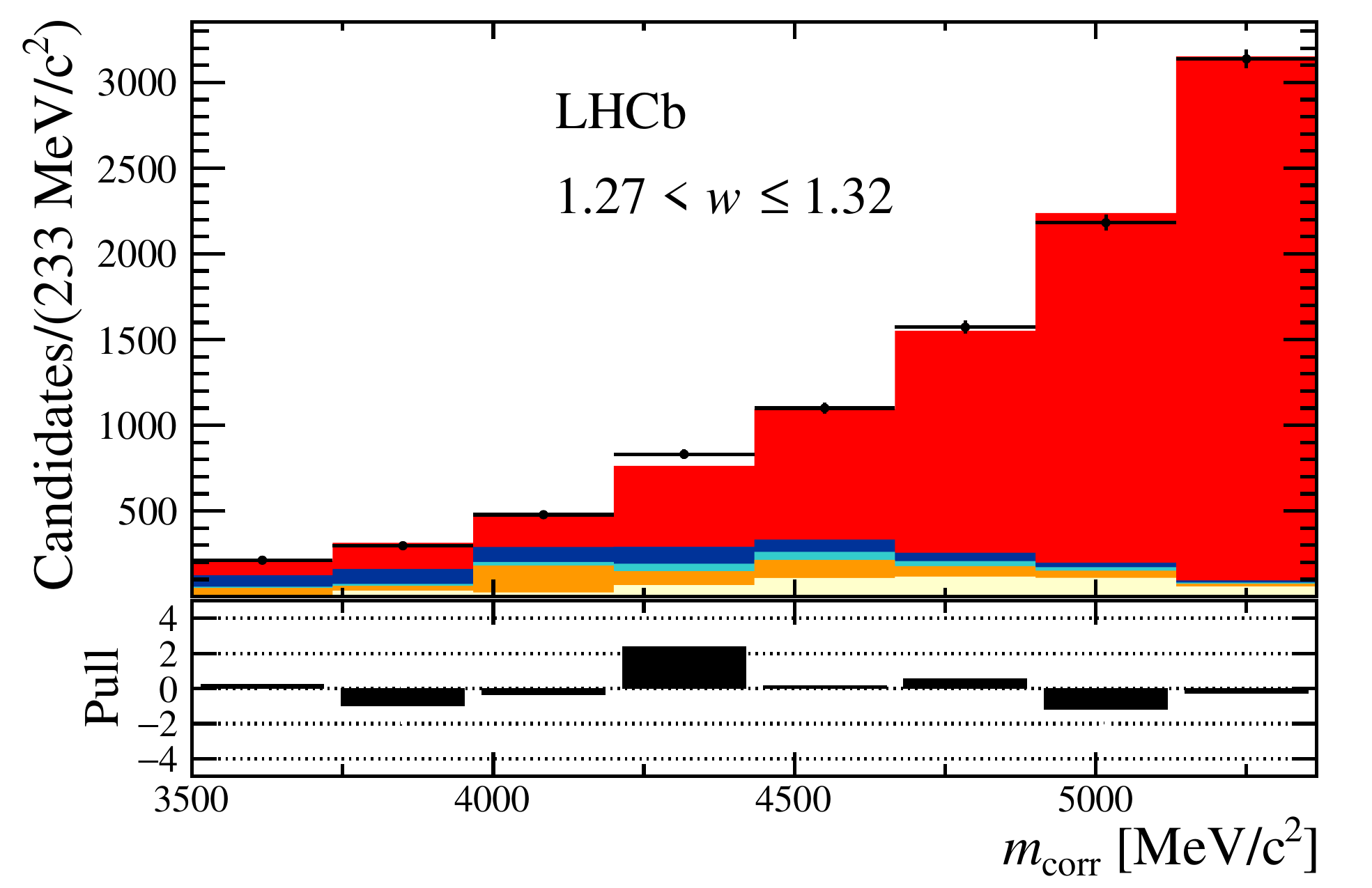}
  \includegraphics[width=0.49\textwidth]{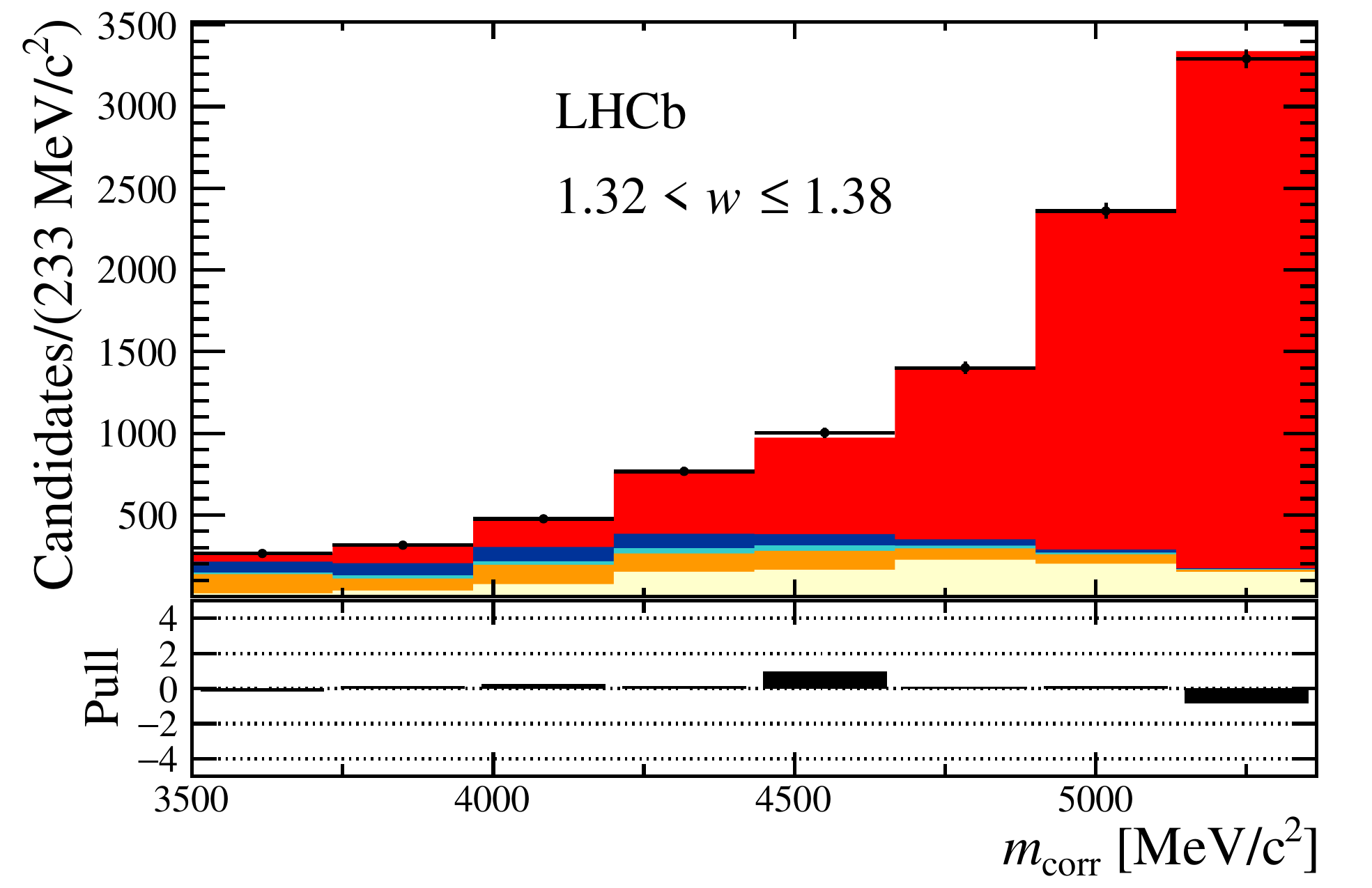}
  \includegraphics[width=0.49\textwidth]{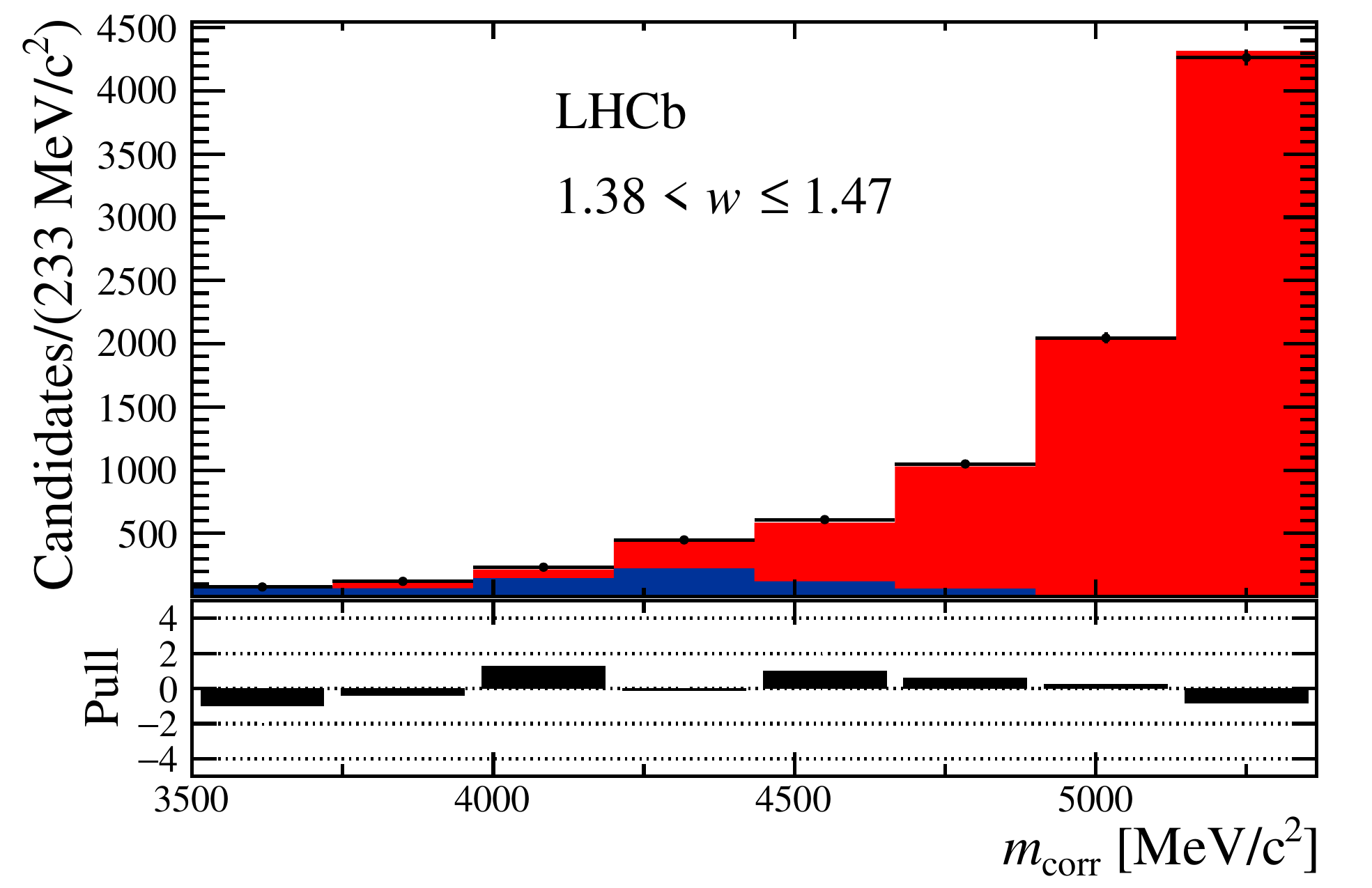}
  \caption{Distribution of the corrected mass, $m_{\rm corr}$, for the seven bins of \w, overlaid with the fit results. The \mbox{\BsToDsonetaunu} and the \mbox{\BsToDsonemunu} components are combined in \mbox{\BsToDsonelnu}. Below each plot, differences between the data and fit are shown, normalised by the uncertainty in the data.}
  \label{fig:yields}
\end{figure}

The free parameters in the fit are the signal yield, the relative abundances of \mbox{\BsToDsstaunu} and \BsToDsonemunu
candidates with respect to that of the signal, and the fraction of combinatorial background. The total fraction of backgrounds from 
\HcTomunu decays is fixed to the expected value using the measured branching fractions and selection efficiencies obtained from simulation. A 40\% uncertainty is assigned to this component to account for the uncertainties on the branching fractions~\cite{PDG2019}. The \BsToDsonetaunu
contribution is also fixed assuming a value of its ratio with respect to the muonic mode equal to the SM prediction for $\BR(\BdToDsttaunu)/\BR(\BdToDstmunu)$~\cite{Fajfer:2012vx} under the assumption that this ratio is identical for \Bs meson decays. The contribution of this decay to the fit is negligible.
The Barlow-Beeston ``lite" technique~\cite{Barlow:1993dm,Cranmer:1456844} is applied to account for the limited size of the simulation samples.
The distributions of \mcorr with the fit overlaid are shown in \Figref{fig:yields}.

\begin{figure}[tb]
  \centering
  \includegraphics[width=0.48\textwidth]{Fig3a.pdf}
  \includegraphics[width=0.48\textwidth]{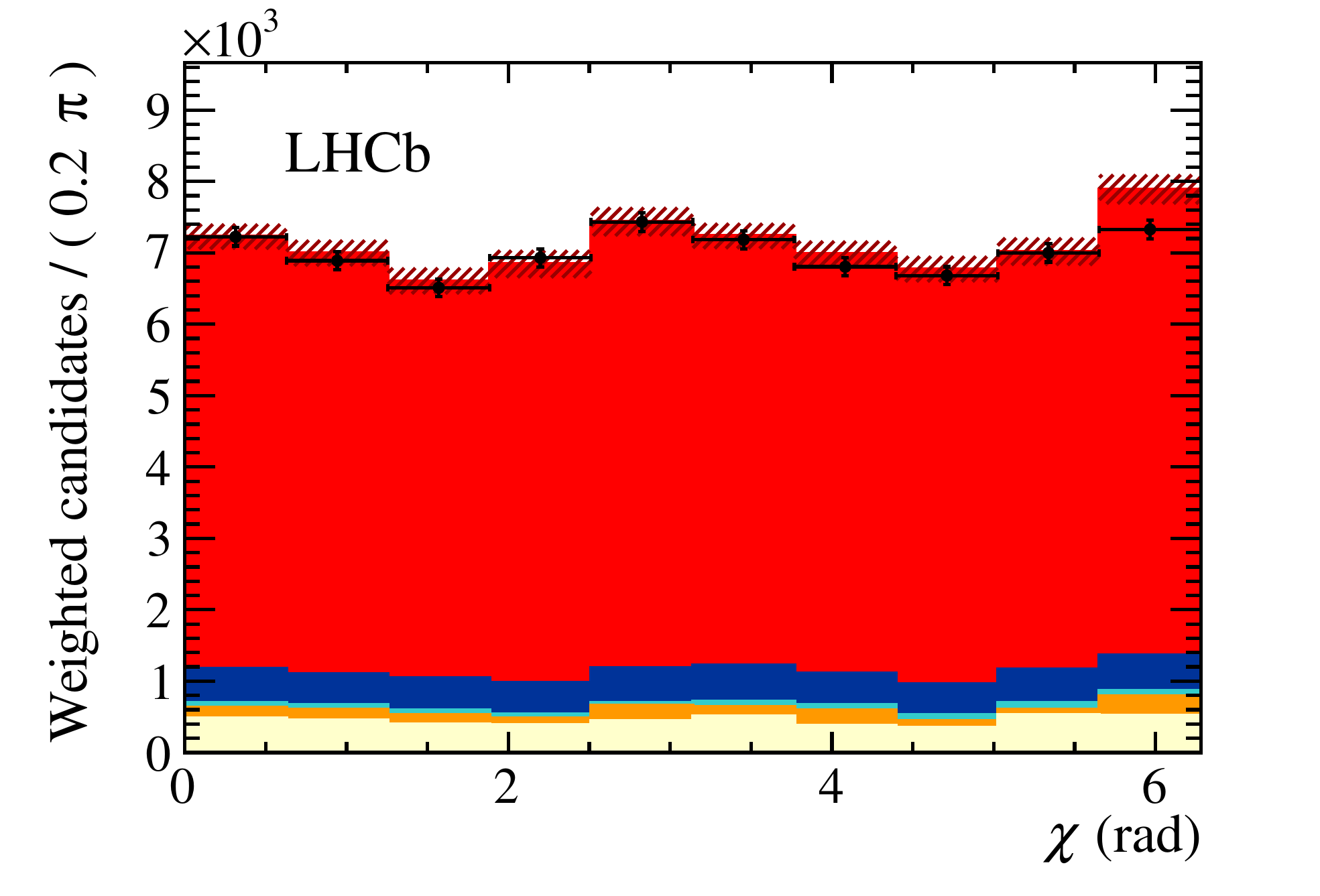}
  \includegraphics[width=0.48\textwidth]{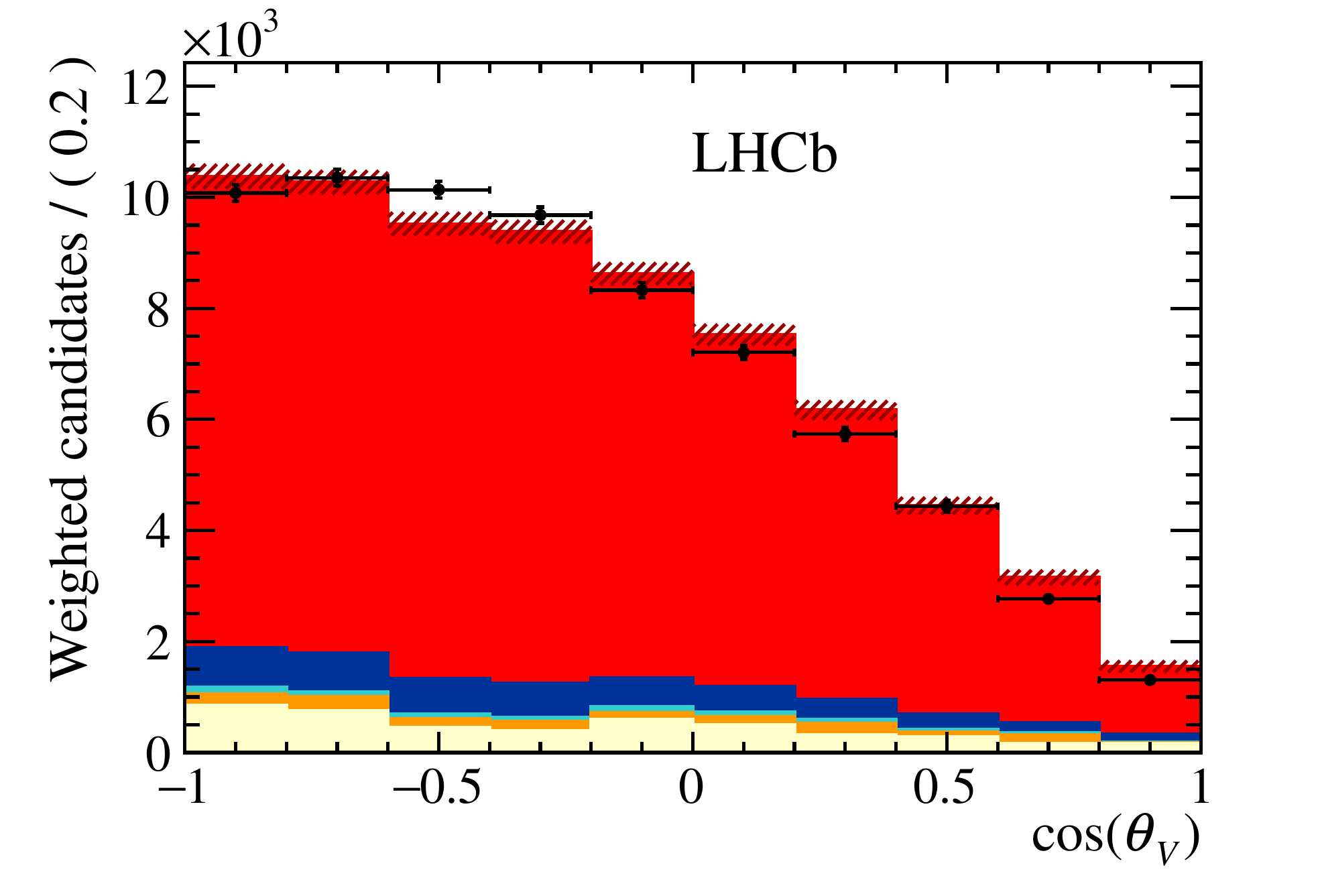}
  \includegraphics[width=0.48\textwidth]{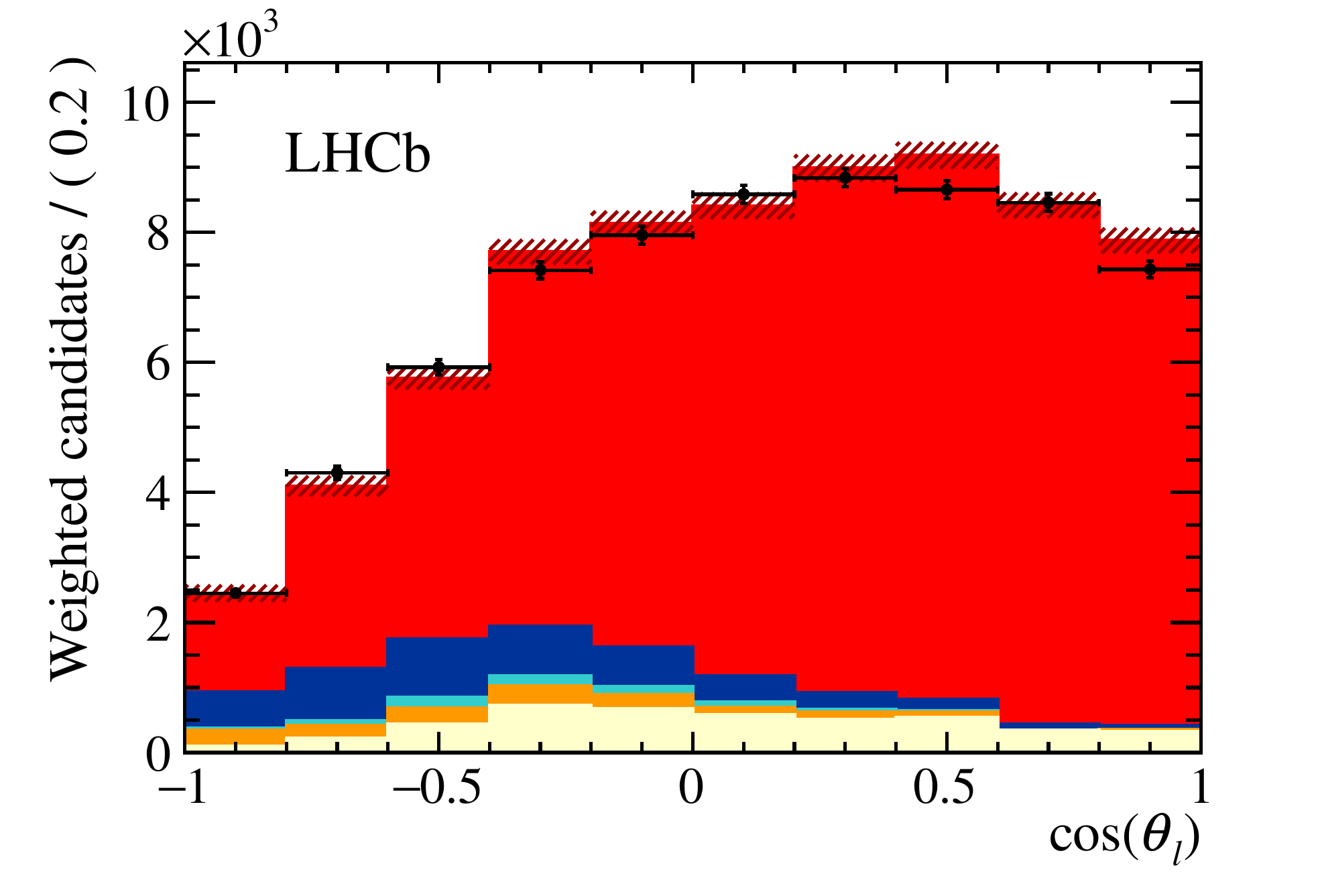}
  \caption{Distribution of (top right) $\chi$, (bottom left) $\cos(\thetaV)$ and (bottom right) $\cos(\thetal)$ integrating over $w$ and the other decay angles from data (black points) compared to the distribution from simulation with their relative size extracted from the fit to the corrected mass. The \BsToDsonetaunu and the \BsToDsonemunu components are combined in \BsToDsonelnu.
  The uncertainties on the templates, indicated by the hashed areas in the figures, are a combination from all templates.}
  \label{fig:dataMCcomparisons}
\end{figure}

Using the fractions obtained from the fit, data and simulated distributions of the angular variables $\cos(\thetal)$, $\cos(\thetaV)$, and $\chi$, as defined in \Secref{sec:eqs}, are shown in \Figref{fig:dataMCcomparisons}.  
All distributions show good agreement between data and simulation, indicating that integrating over the angles does not introduce biases.

%% file: tabs/bins.tex
\begin{table}[tb]
  \caption{Binning scheme used for this measurement. Only the upper bound for each bin is presented. The lower bound on the first bin corresponds to $w=1$.}
  \label{tabs:bins}
  \begin{adjustbox}{center}
  \bgroup
  \def\arraystretch{1.1}
    \begin{tabular}{lccccccc}
    \toprule
    bin   & 1 & 2 & 3 & 4 & 5 & 6 & 7 \\
    \midrule
    \w & 1.1087 & 1.1688 & 1.2212 & 1.2717 & 1.3226 & 1.3814 & 1.4667 \\
    %\qsq [\gevgevcccc] & 8.1--11.0 & 6.8--8.1 & 5.6--6.8 & 4.4--5.6 & 3.3--4.4 & 2.0--3.3 & 0.0--2.0 \\
    \bottomrule
  \end{tabular}
  \egroup
  \end{adjustbox}
\end{table}

%% file: efficiency.tex
\section{Efficiency correction}
\label{sec:eff}

This analysis requires a precise measurement of all contributions to the efficiency as a function of the true value of the hadronic recoil \wtrue extracted from simulation. However, the overall normalisation of the efficiency is not determined as only its dependency with \wtrue is relevant. 

The total efficiency is 
the product of the geometrical acceptance of the detector, the efficiency of reconstructing all tracks, the trigger requirements, and the full set of kinematic, PID and background rejection requirements. 
Most of the contributions to the total efficiency are obtained using simulation. 
Only the particle identification and the \Dsm selection efficiencies are derived from data using control samples. The muon and hadron PID efficiencies are taken from large data samples of \mbox{\Jpsimumu} and \DstToDzPi decays, respectively~\cite{LHCb-PUB-2016-021}.
These samples are then used to determine the PID efficiencies in bins of \ptot, \pt and number of tracks in the event.
The \Dsm selection efficiency accounts for selecting the regions in the Dalitz plot, as well as the vetoes described in \Secref{sec:selection}. 
This efficiency is determined from a sample of fully reconstructed \BsToDsPi decays as a function of the \Dsm \pt. The efficiencies extracted from data are convolved with the simulation to obtain their dependency on \wtrue.

The efficiencies derived from simulation are extracted by comparing the generator-level simulation, based
on \pythia~\cite{Sjostrand:2006za, *Sjostrand:2007gs} and \evtgen~\cite{Lange:2001uf}, 
to the final reconstructed and selected simulation sample used for the
template fit, omitting the particle identification and the \Dsm selection criteria.

%% file: unfolding.tex
\section{Unfolded yields}
\label{sec:unf}
The measured \BsToDssmunu spectrum from \Secref{sec:yield} must be unfolded to account for the resolution on the \w variable, which is 0.07. The unfolding procedure uses a migration matrix determined from simulation, defined as the probability that a candidate generated in bin $j$ of the \wtrue distribution appears in bin $i$ of the $w$ distribution. The unfolded spectrum is then corrected bin-by-bin using the efficiency described in \Secref{sec:eff}. The combination of the migration matrix and the total efficiency, called the response matrix, is shown in App.~\ref{sec:App_FF}.

The unfolding procedure adopted is based on the singular value decomposition (SVD) method~\cite{Hocker:1995kb} using the RooUnfold package implemented in the Root package~\cite{Adye:2011gm}. The SVD method includes a regularisation procedure that depends upon a parameter $k$, ranging between unity and the number of degrees of freedom, seven in this case. Using simulation, the optimal value for $k$ is found to be $k=5$, which minimises the difference between the yield from the unfolding procedure and the expected yield in each bin. 
The final yields, labelled $N_{\rm corr}^{\rm unf}$, are normalised to unity and presented in \Tabref{tabs:NunfCorr}. 

%% file: systematic.tex
\subsection{Systematic uncertainties}
\label{sec:syst}
Systematic uncertainties on $N_{\rm corr}^{\rm unf}$ originate from the fitted \Dssm and \BsToDssmunu yields, and the efficiency corrections. By varying the determination of the unfolded yields, systematic uncertainties are quantified. Since this analysis is sensitive only to the shape of the decay distribution and the absolute normalisation is unknown, every such variation is normalised to unity.
After normalising, the values are compared to those from the default normalised unfolded yields, and from this the uncertainties are extracted. 

The size of the simulated samples, which are very CPU intensive to generate, is the dominating systematic
uncertainty on the unfolded yields. The simulated sample size is accounted for in the fit by applying the Barlow-Beeston ``lite" technique~\cite{Barlow:1993dm,Cranmer:1456844} when determining the signal yield. Its relative contribution to the systematic uncertainty is assessed by not applying this technique and 
comparing the obtained uncertainties.
The uncertainties due to the size of the control samples used to determine the efficiencies and corrections
are obtained by varying each of the efficiency and correction inputs within their uncertainty, repeating
this 1000 times, and taking the spread as the uncertainty on $N_{\rm corr}^{\rm unf}$.

The uncertainty on the SVD unfolding procedure is determined by repeating the regularisation procedure with 
a different regularisation parameter, $k$. The nominal value used is $k=5$, which is changed to $k=4$ and $k=6$, 
and the difference with the nominal value is assigned as the systematic uncertainty. 

Two systematic uncertainties are determined to account for assumptions in the simulation.
Radiative corrections simulated by the \photos package are known to be incomplete~\cite{Golonka:2005pn,Cali:2019nwp}. The difference in $N_{\rm corr}^{\rm unf}$ from simulated samples with and without \photos is evaluated and a third of the difference is 
assigned following Ref.~\cite{Aubert:2008yv}. The efficiency due to the detector acceptance, and thus the shape of the efficiency
correction, may be affected by the form factors in the HQET model used to generate the simulation, 
which are based on the 2016 HFLAV averages~\cite{HFLAV16}. 
This is studied by weighting both the generator level and fully reconstructed simulated samples to the 2019 HFLAV averages~\cite{HFLAV18}: $\rho^2=1.122\pm0.024$, $R_1(1) = 1.270 \pm 0.026$, and $R_2(1) = 0.852 \pm 0.018$, with correlations $\mathrm{corr}[\rho^2,R_1(1)]=-0.824$, $\mathrm{corr}[\rho^2,R_2(1)]=0.566$, and $\mathrm{corr}[R_1(1),R_2(1)]=-0.715$.
The values of each pair are varied within one standard deviation of their mean, taking into account
their correlation. 
The value of $R_0(1)$ is varied by a $20\%$ uncertainty accounting for finite \bquark- and \cquark-quark masses ~\cite{Fajfer:2012vx}.
These variations result in small changes of the total efficiency and the average difference is taken as the uncertainty.

The effect of the \Bs and \g kinematic corrections is assessed by 
changing the kinematic binning schemes in which the corrections are evaluated. The large effect induced by this change has been checked for statistical fluctuations of the calibration samples. The sample is split randomly into two, after which new corrections and $N_{\rm corr}^{\rm unf}$ yields are calculated. No relevant differences between the $N_{\rm corr}^{\rm unf}$ values of these two samples are found in any $w$ bin.
Hence, the systematic uncertainty is based on the change of binning schemes alone.

The corrections to the hardware and software trigger efficiencies applied to the simulated samples depend on the kinematics and PID of the candidates. The systematic uncertainty is evaluated by changing the binning scheme and the PID selection of the control sample.

The systematic uncertainty due to the kinematic dependence of the \Dsm selection efficiency is assessed by extracting the efficiency as a function of \ptot instead of \pt from the \BsToDsPi control sample.

The systematic uncertainty due to the photon background subtraction, performed 
through the \sPlot method with
fits to the \Dssm invariant mass, 
is assessed by implementing the fit with a third-order Chebyshev polynomial for the background description, and repeating the background subtraction process. 

Systematic uncertainties induced by the tracking corrections, detector occupancy and PID efficiencies are found 
to be negligible as they do not affect the corrected mass distribution nor the shape of the efficiency correction.

\input{tabs/yields_syst.tex}
\input{tabs/correlation.tex}

\subsection{Results}
The $N_{\rm corr}^{\rm unf}$ yields and corresponding systematic and statistical uncertainties per $w$ bin are shown in \Tabref{tabs:NunfCorr}. The correlations between the $N_{\rm corr}^{\rm unf}$ yields including statistical and systematic uncertainties are given in \Tabref{tabs:fulcorrmat}, and the covariance matrix is presented in~\Tabref{tabs:fulcovmat} in App.~\ref{sec:App_FF}. The detector response combined with the reconstruction efficiency is presented in App.~\ref{sec:App_FF}. Together these can be used to constrain form-factor parametrisations.

%% file: tabs/yields_syst.tex
\begin{table}[tb]
  \caption{Fraction of the unfolded yields corrected for the global efficiencies, $N^{\mathrm{unf}}_{\mathrm{corr}}$, for each \w bin.
  Also shown in this table is the
  breakdown of the systematic and statistical uncertainties on $N^{\mathrm{unf}}_{\mathrm{corr}}$.
  These are shown as a fraction of the unfolded yield.}
  \label{tabs:NunfCorr}
  \begin{adjustbox}{max width=\textwidth}
  \bgroup
  \def\arraystretch{1.1}
  \begin{tabular}{lccccccc}

    \toprule
    & \multicolumn{7}{c}{\w bin} \\
    \cmidrule[0.5pt]{2-8}
    & 1 & 2 & 3 & 4 & 5 & 6 & 7 \\
    \midrule
    Fraction of $N^{\mathrm{unf}}_{\mathrm{corr},i}$ &0.183 & 0.144 & 0.148 & 0.128 & 0.117 & 0.122 & 0.158 \\
    \midrule[0.75pt]
    %\multicolumn{8}{c}{} \\
    Uncertainties (\%)  & & & & & & & \\
    \midrule
    Simulation sample size       & 3.5 & 3.0 & 2.8 & 3.1 & 3.4 & 3.0 & 3.7 \\
    Sample sizes for effs and corrections           & 3.6 & 3.2 & 3.0 & 2.8 & 2.8 & 2.7 & 2.8 \\
    SVD unfolding regularisation     & 0.5 & 0.5 & 0.1 & 0.7 & 1.2 & 0.0 & 0.5 \\
    Radiative corrections        & 0.1 & 0.2 & 0.1 & 0.3 & 0.4 & 0.2 & 0.2 \\
    Simulation FF parametrisation  & 0.3 & 0.1 & 0.1 & 0.1 & 0.2 & 0.4 & 0.2 \\
    Kinematic corrections          & 2.4 & 1.0 & 1.1 & 0.1 & 0.2 & 0.1 & 0.9 \\
    Hardware-trigger efficiency       & 0.3 & 0.3 & 0.0 & 0.2 & 0.2 & 0.3 & 0.1 \\
    Software-trigger efficiency    & 0.0 & 0.1 & 0.0 & 0.0 & 0.1 & 0.0 & 0.0 \\
    \Dsm selection efficiency     & 0.5 & 0.2 & 0.3 & 0.3 & 0.2 & 0.1 & 0.3 \\
    Photon background subtraction & 0.0 & 2.3 & 0.8 & 2.9 & 2.0 & 0.9 & 0.4 \\
    \midrule
    Total systematic uncertainty & 5.6 & 5.1 & 4.4 & 5.2 & 5.0 & 4.2 & 4.8 \\
    \midrule
    Statistical uncertainty      & 3.4 & 2.9 & 2.7 & 3.1 & 3.2 & 2.9 & 3.4 \\
  \bottomrule
  \end{tabular}
  \egroup
  \end{adjustbox}
\end{table}

%% file: tabs/correlation.tex
\begin{table}[tb]
  \caption{Correlation matrix for the unfolded data set in bins of \w, including both statistical and systematic uncertainties.}
  \label{tabs:fulcorrmat}
  \begin{adjustbox}{center}
  \bgroup
  \def\arraystretch{1.1}
    \begin{tabular}{l@{\hskip 1.cm}ccccccc}
    \toprule
    \w bin &  1 & 2 & 3 & 4 & 5 & 6 & 7  \\
    \midrule
    1 & 1\phantom{.00} &  &  &  &  &  &  \\
    2 & 0.44 & 1\phantom{.00} &  &  &  &  &  \\
    3 & 0.13 & 0.60 & 1\phantom{.00} &  &  &  &  \\
    4 & 0.19 & 0.32 & 0.48 & 1\phantom{.00} &  &  &  \\
    5 & 0.30 & 0.30 & 0.15 & 0.60 & 1\phantom{.00} &  &  \\
    6 & 0.34 & 0.38 & 0.33 & 0.22 & 0.54 & 1\phantom{.00} &  \\
    7 & 0.27 & 0.34 & 0.34 & 0.27 & 0.07 & 0.32 & 1\phantom{.00} \\
    \bottomrule
  \end{tabular}
  \egroup
  \end{adjustbox}
\end{table}

%% file: FF.tex
\section{Form factor fits}
\label{sec:ff}

The yields $N_{\rm corr}^{\rm unf}$ with corresponding correlation matrix presented in \Secref{sec:unf} can be fit using various form-factor parametrisations. Fits using the commonly used CLN and BGL parametrisations, with the assumptions described in \Secref{sec:eqs}, are presented in the following. 

The values of the form-factor parameters are derived from a \chisq fit with
\begin{equation}
    \chi^2 = \sum_{i,j} \left( N^{\mathrm{unf}}_{\mathrm{corr},i} - N_{\mathrm{exp}, i} \right) C_{ij}^{-1}
    \left( N^{\mathrm{unf}}_{\mathrm{corr},j} - N_{\mathrm{exp}, j} \right) . 
\label{eq:chi2}
\end{equation}
In this expression, $N^{\mathrm{unf}}_{\mathrm{corr},i(j)}$ is the normalised, unfolded and efficiency-corrected yield in bin $i(j)$, $N_{\mathrm{exp}, i(j)}$ is the expected yield in bin $i(j)$ obtained from integrating $\deriv\Gamma_{i(j)}/\deriv w$ from the CLN or BGL parametrisation over the bin, 
and $C_{ij}$ is the covariance matrix describing the statistical uncertainties
from the yields and efficiency corrections.  This \chisq function is minimised for the
CLN and BGL parametrisations separately. For the CLN parametrisation, the fitted value is $\rho^2=1.16\pm0.05$, where the uncertainty is only statistical in nature.

For the BGL parametrisation, the unitarity constraint is considered in the minimisation by adding a Gaussian penalty function~\cite{Bordone:2019vic} to the \chisq defined in \Eqref{eq:chi2}. 
This function is of the form 
\begin{equation}
\theta(U-1)\Bigg(\frac{U-1}{\sigma}\Bigg)^2,
\end{equation} 
where $\theta$ is the Heaviside function, $U$ is the unitarity constraint 
$\sum_{n=0}^{2} (a_n^f)^2 + \sum_{n=0}^{2} (a_n^{\mathcal{F}_1})^2$, and $\sigma$ is the theoretical uncertainty associated with the bound~\cite{Bigi:2016mdz}.
The correlation between the external parameter $a_0^f$ and the fitted parameters $a_1^f$ and $a_2^f$ is not considered due to its small uncertainty. To assess the impact of this choice, the value of $a_0^f$ has been increased (decreased) by $+1(-1)\sigma$. The change in the fitted parameters is observed to be negligible compared with the overall systematic uncertainty which covers for it for any value of the correlation between $a_0^f$ and the rest of the parameters. This can be explained as $a_0^f$ only enters as a nuisance parameter in the unitarity bounds, which is the only source of correlation between these parameters. As the two scaled parameters $a_1^f/a_0^f$ and $a_2^f/a_0^f$ can be much larger than one (as shown in \Figref{fig:contour}) the average magnitude of the correlation diminishes.
%This function has to resemble a barrier that adds nothing to the \chisq when the unitarity constraint is satisfied, and infinite when it is not. As the minimisation is performed numerically, the penalty function has to be continuous and differentiable, so a power law function is chosen. The value of the exponent has to be sufficiently large to ensure a rapid increase outside of the allowed region, but small enough not to induce numerical errors in the minimisation. In this case, the maximum power that satisfies both requirements is 34. The chosen penalty function is then $\left[\sum_{n=0}^{2} (a_n^f)^2 + \sum_{n=0}^{2} (a_n^{\mathcal{F}_1})^2\right]^{34}$. 

The fitted values are 
\mbox{$a_1^f=-0.005 \pm 0.034$}, and \mbox{$a_2^f=1.00^{+0.00}_{-0.19}$}, where the uncertainties are only statistical in nature. 

\subsection{Systematic uncertainties}

The systematic uncertainties on the parameters $\rho^2$, 
$a_1^f$ and $a_2^f$ originate from the same sources as those described in \Secref{sec:syst}. Additional systematic uncertainties originate from the external parameters used in the form-factor fits. A summary of all systematic uncertainties for $\rho^2$, $a_1^f$ and $a_2^f$ is shown in \Tabref{tabs:syst:contribution_unfolded}.

The impact of changes in signal yields or efficiencies has been assessed by repeating the fit with different conditions
and comparing the obtained values to the nominal ones.
In the \chisq fit, the parameters $R_1(1)$ and $R_2(1)$ are fixed to the HFLAV averages~\cite{HFLAV18}. 
The uncertainties on these values are propagated to the CLN fit outcome by changing $R_1(1)$ and $R_2(1)$ within one standard deviation from their average, while accounting for the correlation between these values.
For the BGL fit, the values of the external parameters of the $f(z)$, $g(z)$ and $\mathcal{F}_1(z)$ functions are varied simultaneously within their uncertainty. When the uncertainties are asymmetric the largest is chosen. This process is repeated 1000 times applying the unitarity constraint and the difference between the average of the variations and the nominal value is assigned as a systematic uncertainty.

\input{tabs/breakdown}

\subsection{Results}
An analysis to extract the leading parameters of the form factor describing the semileptonic transition \BsToDssmunu has been performed. Using the CLN parametrisation the result obtained is
\begin{equation*}
\rho^2=1.16\pm0.05\stat\pm0.07\syst,
\end{equation*}
where the mass of the muon has not been neglected. To compare with other published results, the fit is repeated assuming a massless muon, resulting in a small shift of the central value of the $\rho^2$ parameter of about 1.5\%, as shown in \Tabref{tab:FFresults}.
The world-average value of $\rho^2$ for the equivalent \Bd$\rightarrow$\Dstarp\mun\neum decay is $\rho^2=1.122\pm0.015\stat\pm0.019\syst$~\cite{HFLAV18}.
Both values of $\rho^2$ are consistent within their uncertainties.
The measurement is also in agreement with the value obtained in Ref.~{\cite{LHCb-PAPER-2019-041}}, $\rho^2=1.23\pm0.17\stat\pm0.05\syst\pm0.01{\aunit{(ext)}\xspace}$, where the last uncertainty comes from external inputs. That analysis uses $\Bs\rightarrow \D_s^{*-}\mup\nu_\mu$ decays from an independent data set, and where the photon from the \Dssm decay is not reconstructed. 
A comparison with the normalised $\Delta\Gamma/\Delta\w$ spectra inferred from the CLN and BGL parametrisations in Ref.~{\cite{LHCb-PAPER-2019-041}} gives consistent results with the measured \w spectrum in this paper, as shown in  App.~\ref{sec:App_CompVcb}.

\begin{figure}[t]
  \centering
  \includegraphics[width=0.8\textwidth]{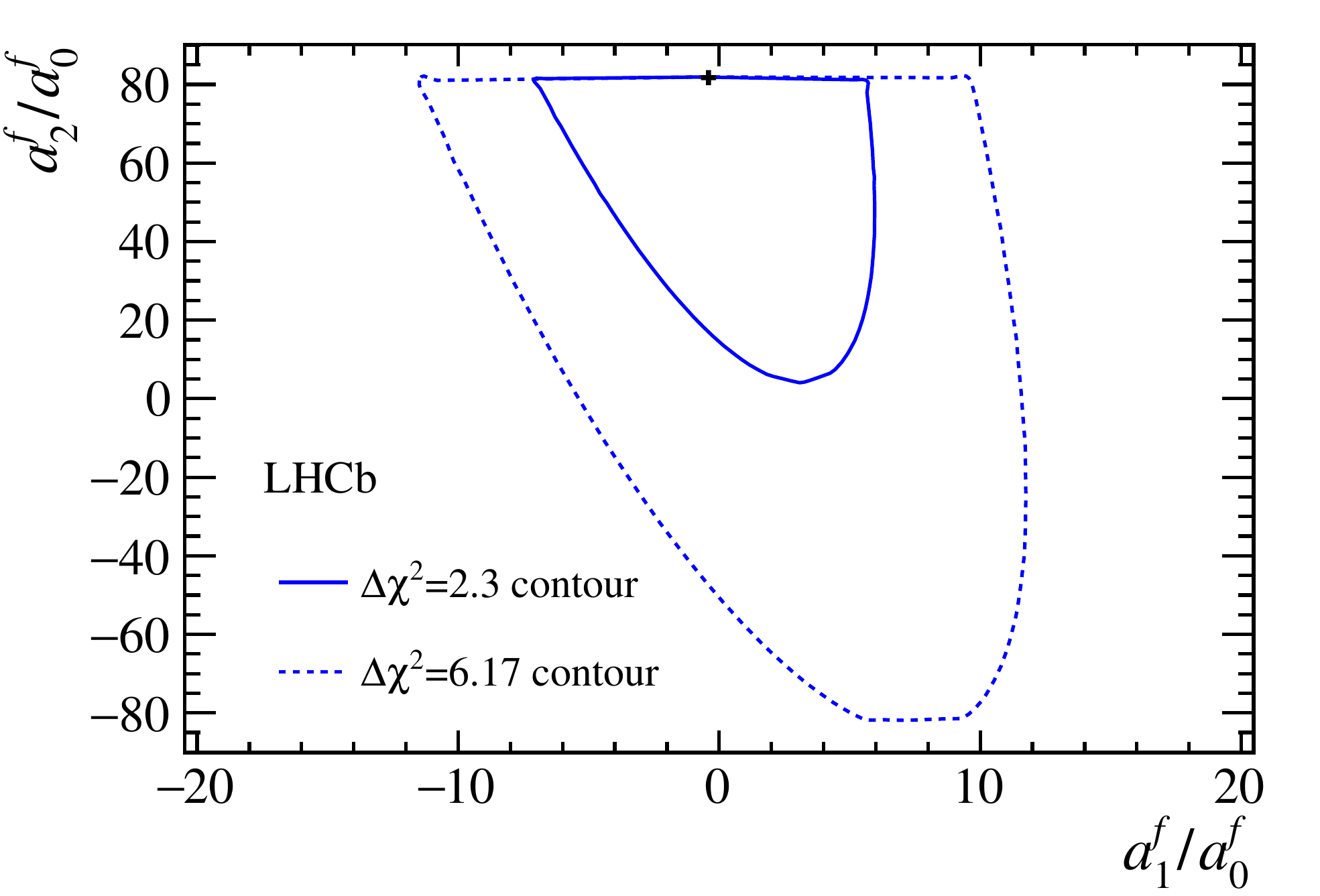}
  \caption{$\Delta$\chisq contours for the scaled parameters $a_1^f/a_0^f$ versus $a_2^f/a_0^f$. The black cross marks the best-fit central value. The solid (dashed) contour encloses the
  $\Delta$\chisq = 2.3 (6.17)
  region.
  The observed shape is due to
  the applied unitarity condition, see \Eqref{eq:unitarity}. 
}
  \label{fig:contour}
\end{figure}

\begin{figure}[ht!]
  \centering
  \includegraphics[width=0.8\textwidth]{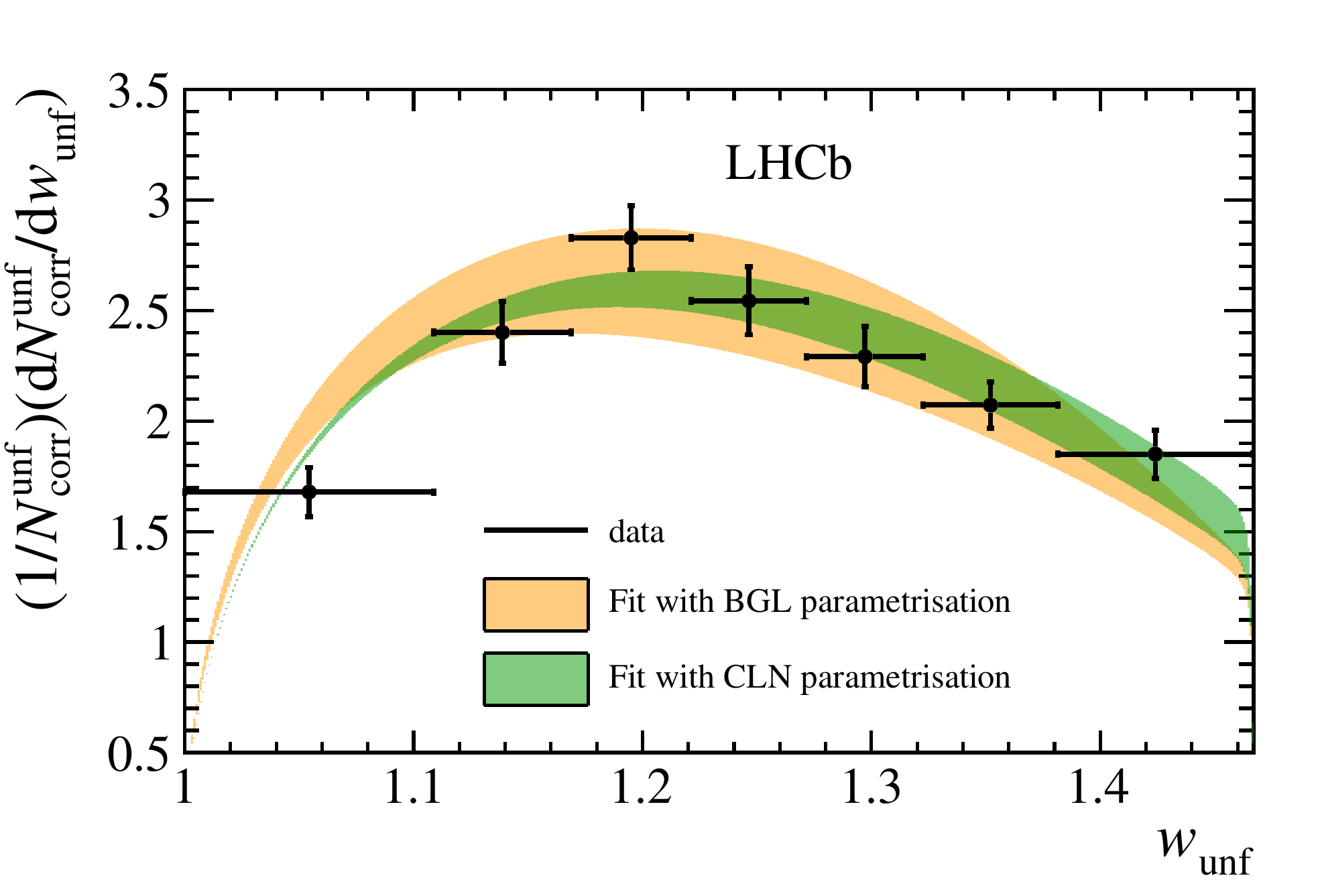}
  \caption{Unfolded normalised differential decay rate with the fit superimposed for the CLN parametrisation (green), and BGL (red).
  The band in the fit results includes both the statistical and systematic uncertainty on the data yields.}
  \label{fig:w_fit}
\end{figure}

Using the BGL parametrisation, the results obtained are
\begin{align*}
a_1^f&=-0.005\pm0.034\stat\pm0.046\syst, \\
a_2^f&=\phantom{-}1.00^{\,+\,0.00}_{\,-\,0.19}\stat ^{\,+\,0.00}_{\,-\,0.38}\syst.
\end{align*}
In \Figref{fig:contour}, the $\Delta$\chisq contours for the scaled parameters $a_1^f/a_0^f$ versus $a_2^f/a_0^f$ are shown; \Figref{fig:contour2} in App.~\ref{sec:inputsBGL} shows the contours of the unscaled $a_1^f$ versus $a_2^f$ parameters. The unitarity constraint results in a non-gaussian distribution of the uncertainty on the $a_2^f/a_0^f$ parameter. 
The fits to the differential decay rate using both parametrisations are shown in \Figref{fig:w_fit}. The $p$-values are $8.2\%$ and $1.3\%$ for the CLN and BGL parametrisations, respectively. The low $p$-values are found to be caused by the third bin in $w$, which is higher than expected for both parametrisations. When artificially decreasing the central value of this bin by one standard deviation, the $p$-values increase to $69.7\%$ and $8.3\%$ for the CLN and BGL parametrisations, respectively. The low $p$-value for the latter fit is explained by the fact that the minimum of the \chisq function without the unitarity constraint lies in the region excluded by this constraint.

The prediction of the decay rate can also be transformed to a prediction of the expected normalised event yields taking into account the efficiency and resolution, which then is fit to the experimental spectrum. Both procedures provide similar results with small differences induced by slightly different bin-by-bin correlations shown in \Tabref{tab:FFresults}. 

\input{tabs/FFresults.tex}

%% file: tabs/breakdown.tex
\begin{table}[tb]
  \caption{Summary of the systematic and statistical uncertainties on the parameters $\rho^2$, $a_1^f$ and $a_2^f$ from the unfolded CLN and BGL fits. The total systematic uncertainty is obtained by adding the individual components in quadrature.}
  \label{tabs:syst:contribution_unfolded}
  \begin{adjustbox}{center}
  \begin{tabular}{l@{\hskip 1.5cm}c@{\hskip 1.5cm}c@{\hskip 1.5cm}c}
    \toprule
    Source                        & $\sigma(\rho^2)$ & $\sigma(a_1^f)$  & $\sigma(a_2^f)$\\
    \midrule                                                           
    Simulation sample size        & 0.053            & 0.036 & $^{\,+\,0.00}_{\,-\,0.35}$ \\
    Sample sizes for efficiencies and corrections           & 0.020            & 0.016 & $^{\,+\,0.00}_{\,-\,0.15}$ \\
    SVD unfolding regularisation  & 0.008            & 0.004 & \text{--} \\
    Radiative corrections         & 0.004            & \text{--} & \text{--} \\ 
    Simulation FF parametrisation & 0.007            & 0.005 & \text{--} \\
    Kinematic corrections         & 0.024            & 0.012 & \text{--} \\
    Hardware-trigger efficiency   & 0.001            & 0.008 & \text{--} \\
    Software-trigger efficiency   & 0.004            & 0.002 & \text{--} \\
    \Dsm selection efficiency     & \text{--}            & 0.008 & \text{--} \\
    Photon background subtraction & 0.002            & 0.015 & \text{--} \\
%    $R_1(1)$ and $R_2(1)$ in fit  & 0.023            & -- & -- \\
%    $R_0(1)$ in fit               & 0.007            & -- & -- \\
%    $a_0^f$  in fit               & -- & 0.000 & 0.00 \\
    External parameters in fit    & 0.024            & 0.002 & $^{\,+\,0.00}_{\,-\,0.04}$ \\
    \midrule                                                        
    Total systematic uncertainty  & 0.068            & 0.046 & $^{\,+\,0.00}_{\,-\,0.38}$\\
     \midrule                                                        
    Statistical uncertainty       & 0.052            & 0.034 & $^{\,+\,0.00}_{\,-\,0.19}$ \\
  \bottomrule
  \end{tabular}
  \end{adjustbox}
\end{table}

%% file: tabs/FFresults.tex
\begin{table}[tb]
\begin{center}
\caption{Results from different fit configurations, where the first uncertainty is statistical and the second systematic.
\label{tab:FFresults}}
\bgroup
\def\arraystretch{1.5}
\begin{tabular}{l l}
\toprule
CLN fit & \\
\midrule
Unfolded fit & $\rho^2=1.16\pm0.05\pm0.07$  \\
Unfolded fit with massless leptons & $\rho^2=1.17\pm0.05\pm0.07$ \\
Folded fit & $\rho^2=1.14\pm0.04\pm0.07$  \\
\midrule[1pt]
BGL fit & \\
\midrule
\multirow{2}{*}{Unfolded fit} & $a_1^f=-0.005\pm0.034\pm0.046$ \\ & $a_2^f=1.00 \kern-0.3em\phantom{.}^{\,+\,0.00}_{\,-\,0.19} \kern-0.3em\phantom{.}^{\,+\,0.00}_{\,-\,0.38}$ \\
\multirow{2}{*}{Folded fit} & $a_1^f=0.039\pm0.029\pm0.046$ \\ & $a_2^f=1.00 \kern-0.3em\phantom{.}^{\,+\,0.00}_{\,-\,0.13} \kern-0.3em\phantom{.}^{\,+\,0.00}_{\,-\,0.34}$ \\
\bottomrule
\end{tabular}
\egroup
\end{center}
\end{table}

%% file: conclusions.tex
\section{Conclusions}
\label{sec:conclusions}

In conclusion, this paper presents for the first time the unfolded normalised differential decay rate for \BsToDssmunu decays as a function of the recoil parameter \w. The unfolded spectrum as a function of \w with the systematic uncertainty per bin is given in~\Tabref{tabs:NunfCorr} and the correlations between these bins in~\Tabref{tabs:fulcorrmat}.
This result allows to constrain $\Bs\to\Dssm\mup\neum$ form-factor parametrisations.
The CLN and BGL form-factor parametrisations have been used to fit the measured spectrum with additional input from $\Bd\to\Dstarm\ell^+\nu_\ell$ decays. Both fits give consistent results when compared to data.

%% file: acknowledgements.tex
\section*{Acknowledgements}
%
% These Acknowledgements valid from 3-May-2019
%
\noindent We express our gratitude to our colleagues in the CERN
accelerator departments for the excellent performance of the LHC. We
thank the technical and administrative staff at the LHCb
institutes.
We acknowledge support from CERN and from the national agencies:
CAPES, CNPq, FAPERJ and FINEP (Brazil); 
MOST and NSFC (China); 
CNRS/IN2P3 (France); 
BMBF, DFG and MPG (Germany); 
INFN (Italy); 
NWO (Netherlands); 
MNiSW and NCN (Poland); 
MEN/IFA (Romania); 
MSHE (Russia); 
MinECo (Spain); 
SNSF and SER (Switzerland); 
NASU (Ukraine); 
STFC (United Kingdom); 
DOE NP and NSF (USA).
We acknowledge the computing resources that are provided by CERN, IN2P3
(France), KIT and DESY (Germany), INFN (Italy), SURF (Netherlands),
PIC (Spain), GridPP (United Kingdom), RRCKI and Yandex
LLC (Russia), CSCS (Switzerland), IFIN-HH (Romania), CBPF (Brazil),
PL-GRID (Poland) and OSC (USA).
We are indebted to the communities behind the multiple open-source
software packages on which we depend.
Individual groups or members have received support from
AvH Foundation (Germany);
EPLANET, Marie Sk\l{}odowska-Curie Actions and ERC (European Union);
ANR, Labex P2IO and OCEVU, and R\'{e}gion Auvergne-Rh\^{o}ne-Alpes (France);
Key Research Program of Frontier Sciences of CAS, CAS PIFI, and the Thousand Talents Program (China);
RFBR, RSF and Yandex LLC (Russia);
GVA, XuntaGal and GENCAT (Spain);
the Royal Society
and the Leverhulme Trust (United Kingdom).

%% file: appendix.tex
\section*{Appendices}
\appendix 

\section{Fitted yields and efficiency}
\label{sec:App_yields}

\begin{figure}[ht!]
  \centering
  \includegraphics[width=0.8\textwidth]{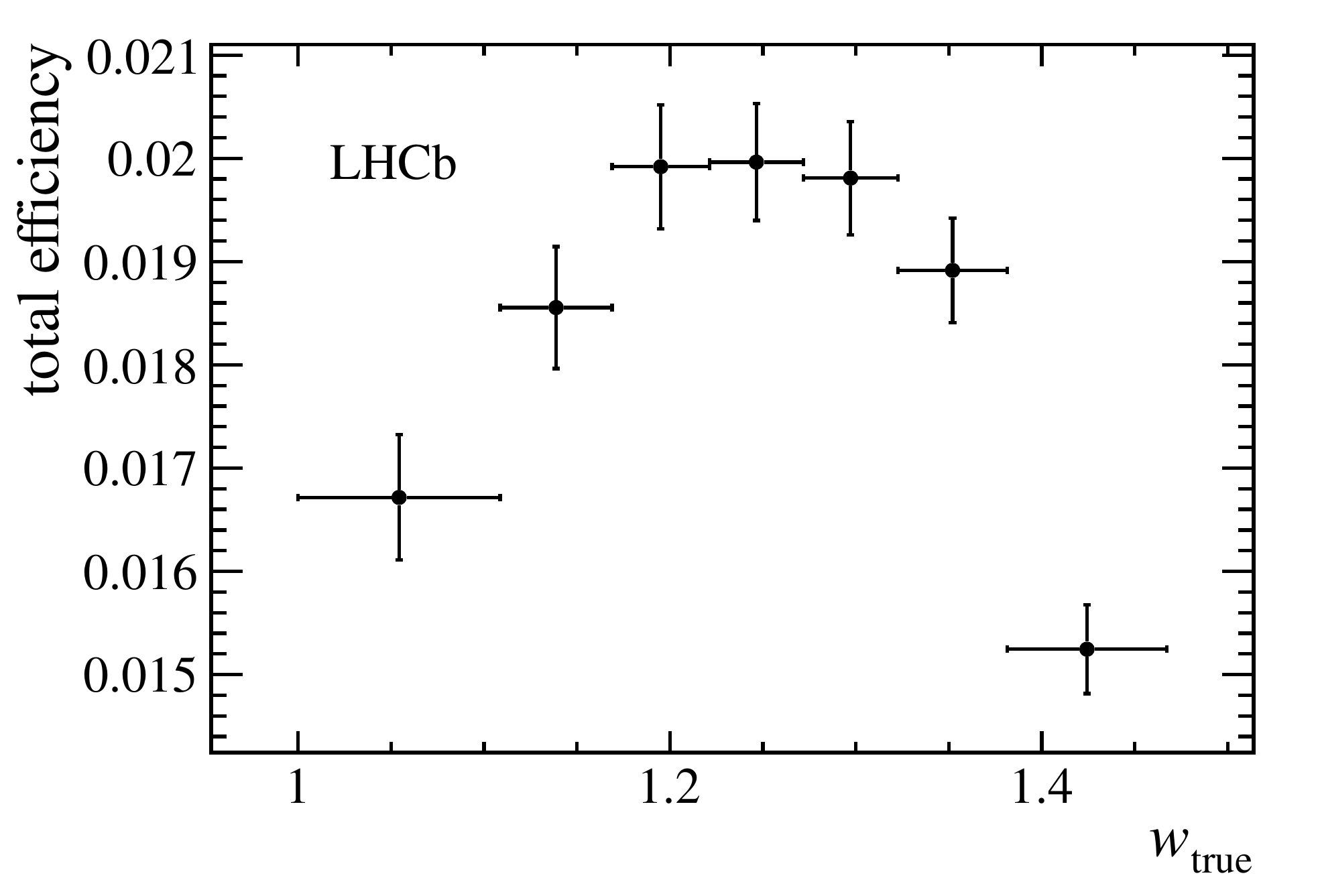}
  \caption{Total efficiency as a function of \wtrue, including the acceptance of the LHCb detector as well as the reconstruction and selection efficiencies.}
  \label{fig:eff}
\end{figure}

Figure~\ref{fig:eff} shows the total efficiency applied to the unfolded signal yields, as a function of \wtrue. It is the combination of the reconstruction and selection efficiencies, including the acceptance of the LHCb detector.

\section{Covariance and response matrices}
\label{sec:App_FF}
This section contains the information needed to reproduce a form-factor fit. To perform the fit using the unfolded, efficiency-corrected and normalised yields given in \Tabref{tabs:NunfCorr}, the corresponding covariance matrix with the combined statistical uncertainties is given in \Tabref{tabs:fulcovmat}.

To transform theoretical predictions into expected signal yields, the response matrix, given in \Tabref{tabs:resp} is needed. This contains the migration matrix (from the true value of $w$ to the reconstructed one) combined with the reconstruction efficiency. The migration matrix is normalised such that the entries within a given bin of $w$ sum up to unity. The absolute efficiencies have not been measured for this analysis.

\input{tabs/covariance}
\input{tabs/response.tex}

\clearpage

\section{Additional information BGL fit}
\label{sec:inputsBGL}
Table \ref{tab:BGLinputs} gives an overview of the fit inputs for the BGL fit.
\input{tabs/BGLinputs.tex}

\begin{figure}[b]
  \centering
  \includegraphics[width=0.8\textwidth]{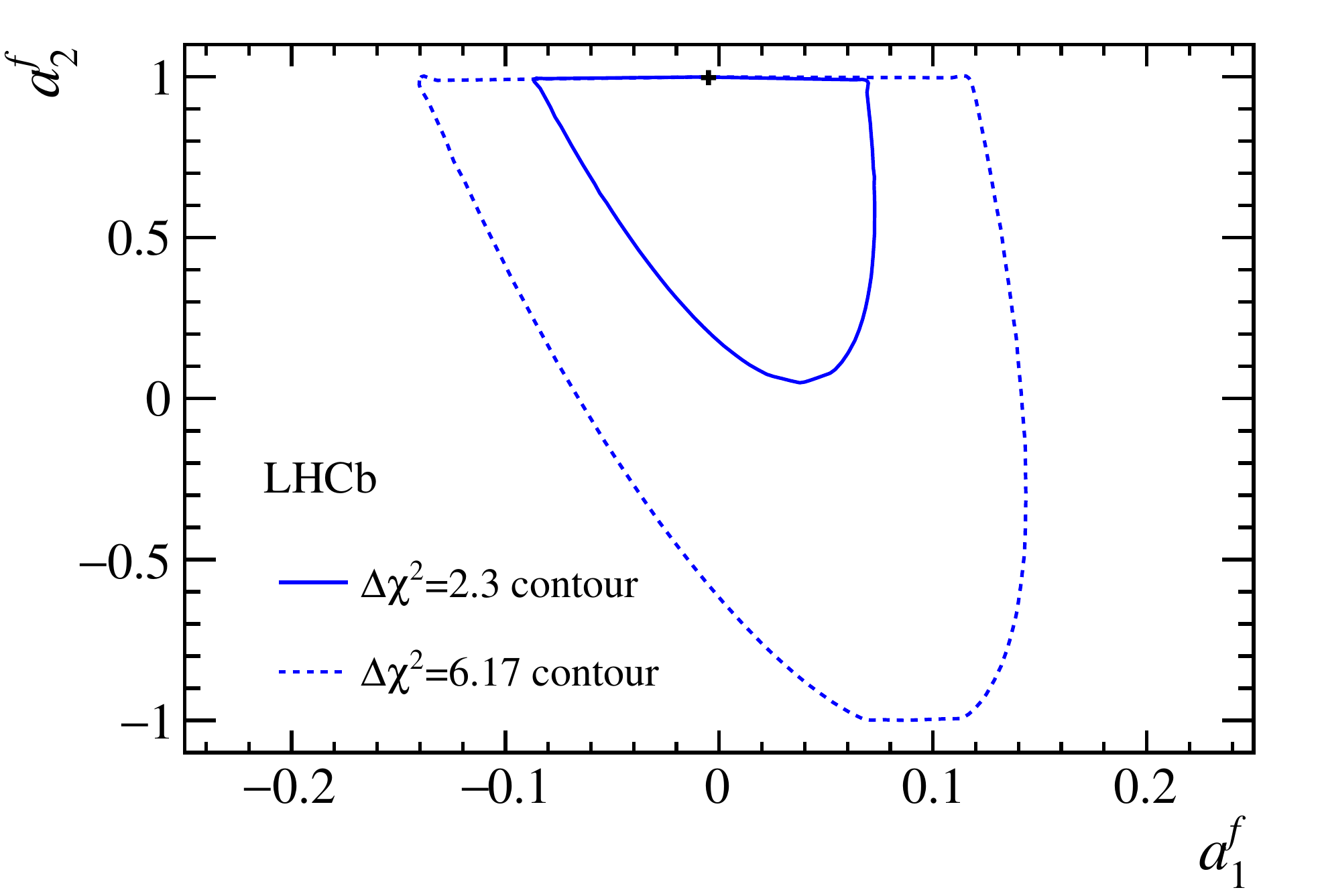}
  \caption{$\Delta$\chisq contours for the scaled parameters $a_1^f$ versus $a_2^f$. The black cross marks the best-fit central value. The solid (dashed) contour encloses the
  $\Delta$\chisq = 2.3 (6.17)
  region.
  The observed shape is due to
  the applied unitarity condition, see Eq. \eqref{eq:unitarity}. 
}
  \label{fig:contour2}
\end{figure}

Figure~\ref{fig:contour2} shows the unscaled $a_1^f$ versus $a_2^f$ contours, equivalent to \Figref{fig:contour}.

\clearpage
\section{Comparison with Phys. Rev. D101 (2020) 072004}
\label{sec:App_CompVcb}
\begin{figure}[ht!]
  \centering
  \includegraphics[width=0.8\textwidth]{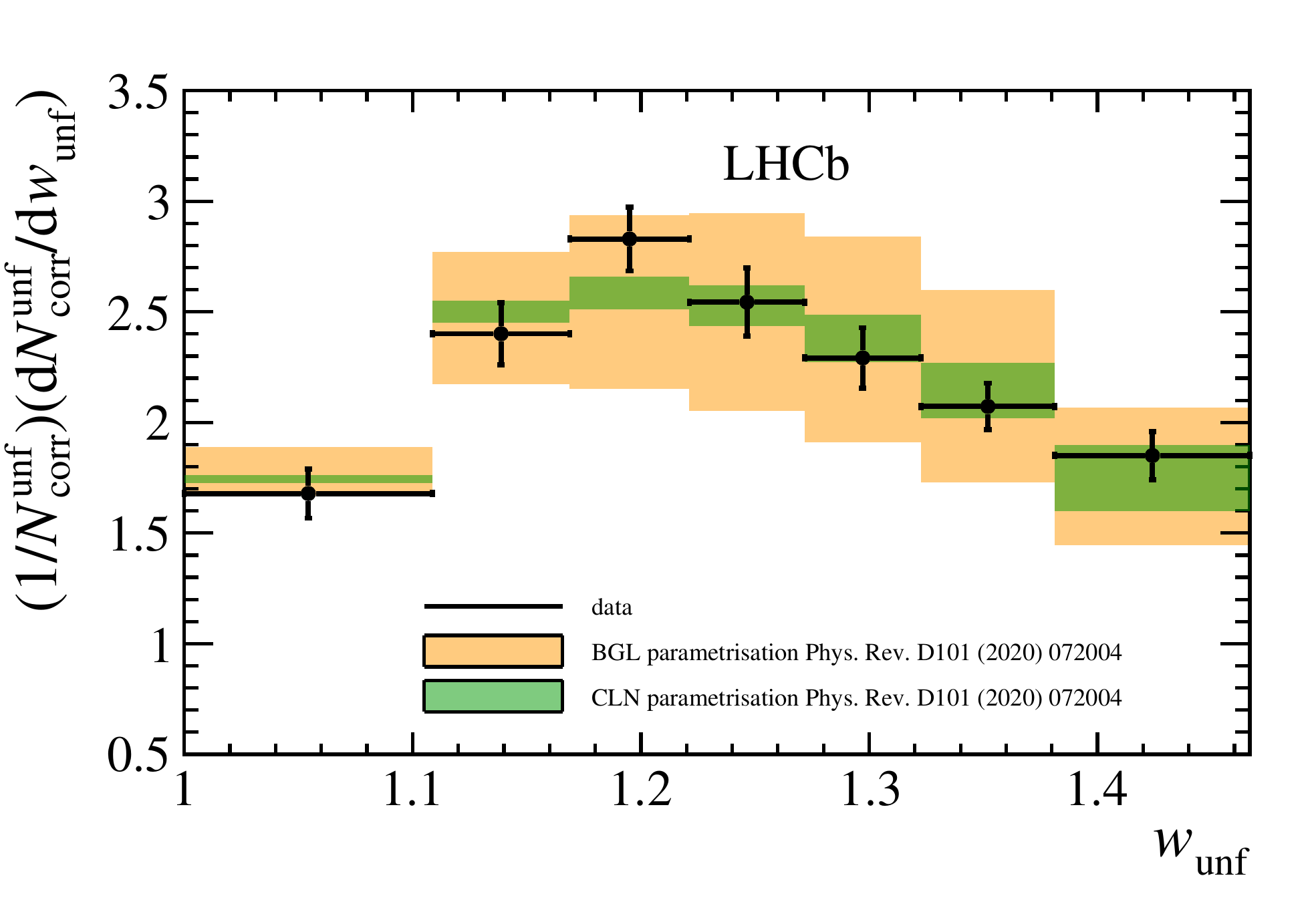}
  \caption{Comparison between the \w spectrum measured in this paper to the normalised $\Delta\Gamma/\Delta\w$ spectra inferred from the CLN and BGL parametrisations in Ref.~\cite{LHCb-PAPER-2019-041}.}
  \label{fig:compwithVcbanalysis}
\end{figure}

The \w spectrum measured in this analysis can be compared with the results obtained in Ref.~\cite{LHCb-PAPER-2019-041} where the form-factor parameters of the \BsToDssmunu decay are measured using a version of the CLN and BGL parametrisations. From this, the normalised $\Delta\Gamma/\Delta\w$ spectrum can be inferred, which is shown in \Figref{fig:compwithVcbanalysis}. The spectrum measured in this paper is consistent with the normalised spectra inferred from both CLN and BGL parametrisations used in Ref.~\cite{LHCb-PAPER-2019-041}.

%% file: tabs/covariance.tex
\begin{table}[tb]
  \caption{Covariance matrix for the unfolded data set in bins of \w, including both statistical and systematic uncertainties in units of $10^{-5}$.}
  \label{tabs:fulcovmat}
  \begin{adjustbox}{center}
  \bgroup
  \def\arraystretch{1.1}
    \begin{tabular}{l@{\hskip 1.cm}ccccccc}
    \toprule
    \w bin [$10^{-5}$] &  1 & 2 & 3 & 4 & 5 & 6 & 7  \\
    \midrule
    1 & \kern-0.5em 16.10 &  &  &  &  &  &  \\
    2 & 4.73 & 7.05 &  &  &  &  &  \\
    3 & 1.21 & 3.81 & 5.63 &  &  &  &  \\
    4 & 1.87 & 2.12 & 2.81 & 6.10 &  &  &  \\
    5 & 2.74 & 1.80 & 0.78 & 3.37 & 5.12 &  &  \\
    6 & 2.42 & 1.82 & 1.38 & 0.98 & 2.17 & 3.19 &  \\
    7 & 3.24 & 2.69 & 2.43 & 2.02 & 0.44 & 1.69 & 8.95 \\
    \bottomrule
  \end{tabular}
  \egroup
  \end{adjustbox}
\end{table}

%% file: tabs/response.tex
\begin{table}[tb]
  \caption{Response matrix, containing the migration from \wtrue to $w$ bins together with the total efficiency in units of $10^{-4}$.}
  \label{tabs:resp}
  \begin{adjustbox}{center}
  \bgroup
  \def\arraystretch{1.1}
    \begin{tabular}{lrrrrrrr}
    \toprule
    $[10^{-4}]$ & \multicolumn{7}{c}{\wtrue} \\
    \cmidrule[0.5pt]{2-8}
    $w$ &  1 & 2 & 3 & 4 & 5 & 6 & 7  \\
    \midrule
    1 & 132.0 & 29.9 & 11.0 & 6.1 & 2.7 & 2.4 & 1.0 \\
    2 & 22.4 & 111.0 & 36.3 & 11.1 & 5.0 & 3.8 & 1.4 \\
    3 & 6.0 & 28.7 & 109.0 & 35.9 & 12.3 & 6.6 & 4.8 \\
    4 & 4.6 & 9.8 & 27.0 & 102.0 & 34.6 & 12.3 & 5.7 \\
    5 & 1.4 & 4.4 & 8.9 & 30.3 & 98.0 & 33.7 & 10.3 \\
    6 & 0.8 & 0.7 & 5.0 & 8.5 & 34.5 & 97.0 & 30.9 \\
    7 & $-0.1$ & 0.7 & 2.2 & 5.7 & 11.0 & 33.5 & 98.5 \\
    \bottomrule
  \end{tabular}
  \egroup
  \end{adjustbox}
\end{table}

%% file: tabs/BGLinputs.tex
\begin{table}[h!]
\begin{center}
\caption{Fit inputs used for the BGL fit, taken from Ref.~\cite{Gambino:2019sif} and Ref.~\cite{Bigi:2017jbd}.
\label{tab:BGLinputs}}
\bgroup
\def\arraystretch{1.5}
\begin{tabular}{c c}
\toprule
BGL parameter           & Value \\
\midrule
$a_0^f$                 & $0.01221 \pm 0.00016$                       \\
%$a_1^f$                & $0.006\left(^{\,+\,32}_{\,-\,45}\right)$    \\
%$a_2^f$                & $ -0.2\left(^{\,+\,12}_{\,-\,8}\right)$     \\
\midrule
$a_1^{\mathcal{F}_1}$   & \kern-1.5em\phantom{\,$^{-}$\,0}$0.0042 \pm 0.0022$   \\
$a_2^{\mathcal{F}_1}$   & \kern-1.5em$-0.069^{\,+\,0.041}_{\,-\,0.037}$   \\
\midrule
$a_0^{g}$               & \kern-0.8em$0.024^{\,+\,0.021}_{\,-\,0.009}$     \\
$a_1^{g}$               & \kern-0.8em$0.05^{\,+\,0.39}_{\,-\,0.72}$     \\
$a_2^{g}$               & \kern-0.8em$1.0^{\,+\,0.0}_{\,-\,2.0}$       \\
\midrule[0.75pt]
$a_0^{\mathcal{F}_2}$   & \kern-1.3em\phantom{-0}$0.0595\pm 0.0093$                 \\
$a_1^{\mathcal{F}_2}$   & \kern-1.3em$-0.318 \pm 0.170$                  \\
\bottomrule
\end{tabular}
\egroup
\end{center}
\end{table}

%% file: LHCb_Authorship_17-Dec-2019.tex
% LHCb collaboration author list
% Data extracted on October 3rd, 2020 at 5:03pm for reference date 17-Dec-2019
\centerline
{\large\bf LHCb collaboration}
\begin
{flushleft}
\small
R.~Aaij$^{31}$,
C.~Abell{\'a}n~Beteta$^{49}$,
T.~Ackernley$^{59}$,
B.~Adeva$^{45}$,
M.~Adinolfi$^{53}$,
H.~Afsharnia$^{9}$,
C.A.~Aidala$^{80}$,
S.~Aiola$^{25}$,
Z.~Ajaltouni$^{9}$,
S.~Akar$^{64}$
% ...

{\footnotesize \it

% ...
$ ^{9}$Universit{\'e} Clermont Auvergne, CNRS/IN2P3, LPC, Clermont-Ferrand, France\\
% ...
$ ^{25}$INFN Sezione di Milano, Milano, Italy\\
% ...
$ ^{31}$Nikhef National Institute for Subatomic Physics, Amsterdam, Netherlands\\
% ...
$ ^{45}$Instituto Galego de F{\'\i}sica de Altas Enerx{\'\i}as (IGFAE), Universidade de Santiago de Compostela, Santiago de Compostela, Spain\\
% ...
$ ^{49}$Physik-Institut, Universit{\"a}t Z{\"u}rich, Z{\"u}rich, Switzerland\\
% ...
$ ^{53}$H.H. Wills Physics Laboratory, University of Bristol, Bristol, United Kingdom\\
% ...
$ ^{59}$Oliver Lodge Laboratory, University of Liverpool, Liverpool, United Kingdom\\
% ...
$ ^{64}$University of Cincinnati, Cincinnati, OH, United States\\
% ...
$ ^{67}$Syracuse University, Syracuse, NY, United States\\
% ...
$ ^{80}$University of Michigan, Ann Arbor, United States, associated to $^{67}$\\
\bigskip
% ...
\medskip
}
\end{flushleft}